\documentclass[journal]{IEEEtran}

\ifCLASSINFOpdf \else \fi

\usepackage{graphicx}

\usepackage[cmex10]{amsmath}
\usepackage{amssymb}

\usepackage{algorithm}
\usepackage{algpseudocode}
\usepackage{array}
\ifCLASSOPTIONcompsoc
 \usepackage[caption=false,font=normalsize,labelfont=sf,textfont=sf]{subfig}
\else
 \usepackage[caption=false,font=footnotesize]{subfig}
\fi
\usepackage{url}
\usepackage[noadjust]{cite}

\usepackage{amssymb}

\newcommand{\Eb}{\boldsymbol{E}}
\newcommand{\Hb}{\boldsymbol{H}}
\newcommand{\Jb}{\boldsymbol{J}}

\newcommand{\Gb}{\mathbf{G}}
\newcommand{\Ib}{\mathbf{I}}
\newcommand{\rb}{\boldsymbol{r}}

\newcommand{\bb}{\boldsymbol{b}}

\newcommand{\eb}{\boldsymbol{e}}

\newcommand{\nb}{\mathbf{n}}
\newcommand{\bm}[1]{\boldsymbol{#1}}
\newcommand{\zhat}{\hat{\mathbf{z}}}

\newcommand{\rhob}{\boldsymbol{\rho}}

\newcommand{\Sb}{\mathbf{S}}
\newcommand{\Zb}{\mathbf{Z}}
\newcommand{\Vb}{\mathbf{V}}
\newcommand{\sign}{\mathrm{sgn}}

\begin{document}

\title{Simulation of S-parameters of general multilayer boxed PCP's with the method of moments and the scattering matrix algorithm}

\author{{$^1$A.~O.~Makarenko, $^1$P.~Zheglova, $^1$R.~Gaponenko, $^2$R.~V.~Salimov, $^2$R.~I.~Tikhonov, and $^1$A.~A.~Shcherbakov}
\thanks{$^1$ITMO University, St-Petersburg, RUSSIA}
\thanks{$^2$LLC GAMMA Tech, St-Petersburg, RUSSIA}}



\maketitle

\begin{abstract}
This paper addresses the problem of modeling high-frequency parameters for shielded printed circuit boards (PCBs) using the Green’s function for a multilayer waveguide and the S-matrix formalism. The proposed approach incorporates a simple, intuitive method for calculating the full tensor Green’s function. The corresponding matrix elements are described using three sets of S-matrices associated with the dielectric layers of the PCB substrate. The method of moments (MoM) is implemented using surface "roof-top" functions, as well as volumetric pulse and linear basis functions; all overlap integrals involving these functions and the waveguide modes are derived in explicit form, enabling the efficient modeling of both planar metallization layers and vertical interconnects. To accelerate computations, a new approach to deriving the expansion of the MoM matrix elements is proposed, which facilitates the application of the fast Fourier transform.
\end{abstract}


\IEEEpeerreviewmaketitle

\section{Introduction}
\IEEEPARstart{M}{ultilayer} printed circuit board (PCB) technology is a key technology in modern microelectronic. Multilayer PCBs are manufactured by depositing thin layers of a conductive material -- typically copper -- between layers of dielectric materials, such as Teflon, quartz, aluminum oxide, or gallium arsenide, and connecting these conductive layers using conductive lines (interconnects). Modern PCBs can have variable and complex geometry, but are characterized by the presence of the following components: a multilayer dielectric substrate consisting of homogeneous layers, which serves to insulate the conductive elements and provide mechanical support for the system; planar elements printed within the substrate at the dielectric layer boundaries; vias or interconnects that pass through the dielectric layers to link the planar elements; and various electrical components, such as resistors, capacitors, inductors, and transistors, attached to conductive pads on the outer surfaces of the PCB, e.g., \cite{Chandler1995, Walter1996, Prentice1997, Morrissey1998, Bai2025, Zeng2025}.

Thus, a distinctive feature of PCBs is their layered dielectric structure, in which, to a good approximation, the dielectric properties vary in only one spatial dimension. This structure can be modelled either as an open system, with layers extending infinitely in the transverse direction, or as a shielded system, essentially a waveguide with a layered structure oriented perpendicular to its main axis. PCBs can be modelled as longitudinally infinite or as having upper and/or lower conductive boundaries; in a shielded structure, the side walls are often assumed to be perfectly conducting. By imposing specific constraints on the dimensions of the structure and ensuring impedance matching at the upper or lower boundaries, the waveguide model can also be adapted to simulate open structures \cite{Rautio2007}.

Presence of thin highly conductive elements makes application of methods involving discretization of volume inefficient. A better approach is to benefit from analytical solutions inherent to layered dielectric structures, and electromagnetic surface integral equations solved by the Method of Moments (MoM) \cite{Gibson2008,Swanson2003-fi,Lancellotti2022,Okhmatovski2024}. MoM is a projection method allowing one to transform an integral equation (IE) into a system of linear equations by expanding unknown currents and known source fields into simple basis and weight functions. In the problems discussed in this paper, the Galerkin method is typically employed, in which the weighting functions coincide with the basis functions. The resulting formulations are often referred to as the 2.5-dimensional (2.5D) method of moments. MoM has been formulated for the electric field integral equation (EFIE) \cite{Khalil1999, Davidson2005, Crespo-Valero2006}, an approach adopted in this paper, the magnetic field integral equation (MFIE) \cite{Davidson2005, Abdul-Gaffoor2000,Abdul-Gaffoor2002}, and the mixed potential or combined field integral equation (MPIE, MFIE) \cite{Michalski1997,Kinayman2005, Okhmatovski2009, Chew2001, YlaOijala2003}. The kernel of the integral equation is the corresponding Green's function for the host medium. The main advantage of the MoM is that the solution is sought only inside the conductors, bypassing the need to solve the electromagnetic problem for the whole dielectric host.

In a shielded PCB, which is considered below, the Green's function for the layered substrate has a form of infinite double series. Convergence of these series can be quite slow when the current source and the field observation point are located at the same of nearby longitudinal coordinates. Since the linear system matrix is dense, the cost of computing the matrix elements and solving the MoM linear equation system can be significant. Various acceleration techniques, of which an overview is given in \cite{Kinayman1995}, have been proposed based on the Kummer transformation \cite{Hashemi1995,Eleftheriades1996a,Khalil1999,Crespo-Valero2006,Nguyen2015}, Ewald's method (more suitable for scalar potential Green's function) \cite{Park1997,Soler2008,delaCruz2023}, and the Fast Fourier Transform (FFT) \cite{Rautio1986,Hill1991,Rautio2014a,Rautio2015,Okhmatovski2024}. FFT-accelerated MoM matrix fill can be combined with FFT-accelerated matrix-vector multiplications \cite{Strang1986, Strang1989} within iterative linear solvers. Fast multipole method \cite{Zhao1998} has also been applied to speed-up iterative solutions. Additionally, symmetry conditions and the area of influence between the basis functions can be taken into account to reduce the problem size \cite{Golestanirad2010,Golestanirad2010a}. 

An important aspect of the mentioned 2.5D MoM formulations  is a necessity to perform a substantial volume of analytical derivations, involving the calculation of the host Green's function components and overlap integrals of the waveguide modes with the MoM basis functions. Several studies have proposed ways to do this for layered structures hosting arbitrary number of layers and flat metallizations parallel to the waveguide layer interfaces. Such configurations require $xx$, $yy$ and $xy$, i.e. transverse Green's function components and overlap integrals of 2D basis functions with the transverse modal functions. MoM analysis of a two-layer dielectric waveguide with a single thin metallization was formulated by \cite{Rautio1987,Rautio2005}. Transmission-line approach for calculation of transverse Green's function components in a layered waveguide with arbitrary number of dielectric layers and thin flat metallizations is proposed in \cite{Melcon1999,Crespo-Valero2006}. This approach is applied in \cite{Stevanovic2009} to modelling waveguide fed patch antennas and arrays, and in \cite{Michalski2019} to modelling metallizations with anisotropy. Generalized scattering matrix approach (GSM) was developed and applied to a problem of a single metallization in a waveguide in \cite{Khalil1999,Khalil1999a}. The GSM approach has been implemented for modelling one strip and one slot discontinuity in a waveguide by \cite{Yakovlev1999,Yakovlev2000}, and generalized to modelling aperture-coupled patch amplifier arrays in \cite{Yakovlev2002}. Three layer Green's function has been explicitly derived through the use of transmission matrices in \cite{Li1997}. \cite{Molina2020} applied a multimode equivalent network (MEN) to modelling flat metallizations in a two and three layer boxed structures. In addition to PCB modelling, applications of the 2.5D MoM methodology for thin flat metallizations in a waveguide include modelling frequency selective surfaces \cite{Ohira2005} and modelling thick irises in a waveguide by introducing a correction to the IE kernel \cite{Stevanovic2004,Stevanovic2006}. A comprehensive unified approach for treating thin flat metallization modelling in a waveguide by the MoM with an accelerated FFT-based fast matrix filling and fast matrix solver is presented in \cite{Rautio2014}. 

A further incorporation of vertical currents requires introduction of vertical surface or volume basis functions, the calculation of the $xz$, $yz$, $zx$, $zy$ and $zz$ components of the Green's function, and evaluation of the corresponding additional overlap integrals. In PCB modelling, vertical Green's function components are needed to model co-axial feed pins, microstrip bends, bridges and T-junctions, shielded subarrays to suppress electromagnetic interference (EMI) \cite{Li2021}, wire-shaped interconnect elements (vias) and metallizations of finite thickness \cite{Rautio2019,Rautio2020,Rautio2021}. 

In a shielded environment, EFIE Green's function kernel components for a two-layer waveguide were proposed in \cite{Nikellis2004,Nikellis2006}, where the authors used a 2D basis function expansion similar to that of \cite{Aroudaki1994}. However, these two studies only show the proposed methodology in a homogeneous medium with a top and bottom conducting cover and one flat metallization in between. A much more general theoretical base for incorporation of the vertical currents with MoM has been developed and implemented by \cite{Rautio2019,Rautio2020,Rautio2021} using surface and volume rooftop basis functions in EFIE formulation. The most comprehensive treatment of this approach, applicable to modelling stacks of thin and thick flat metallic layers, vertical interconnecting structures, bent structures of finite thickness and non-rectangular flat and thick metallizations can be found in \cite{Rautio2020}. In this book, the $z$-dependence is expressed with trigonometric functions, the special cases of two- and three-layered media are considered and a procedure for construction of the Green's function components for multilayered media is described. Additionally, other aspects such as accounting for the metal Ohmic losses and metal surface roughness are considered. This book and other works of its authors are the basis for the commercial software package Sonnet \cite{Sonnet}. 

\begin{figure}[h]
	\centering
	\includegraphics[width=0.47\textwidth]{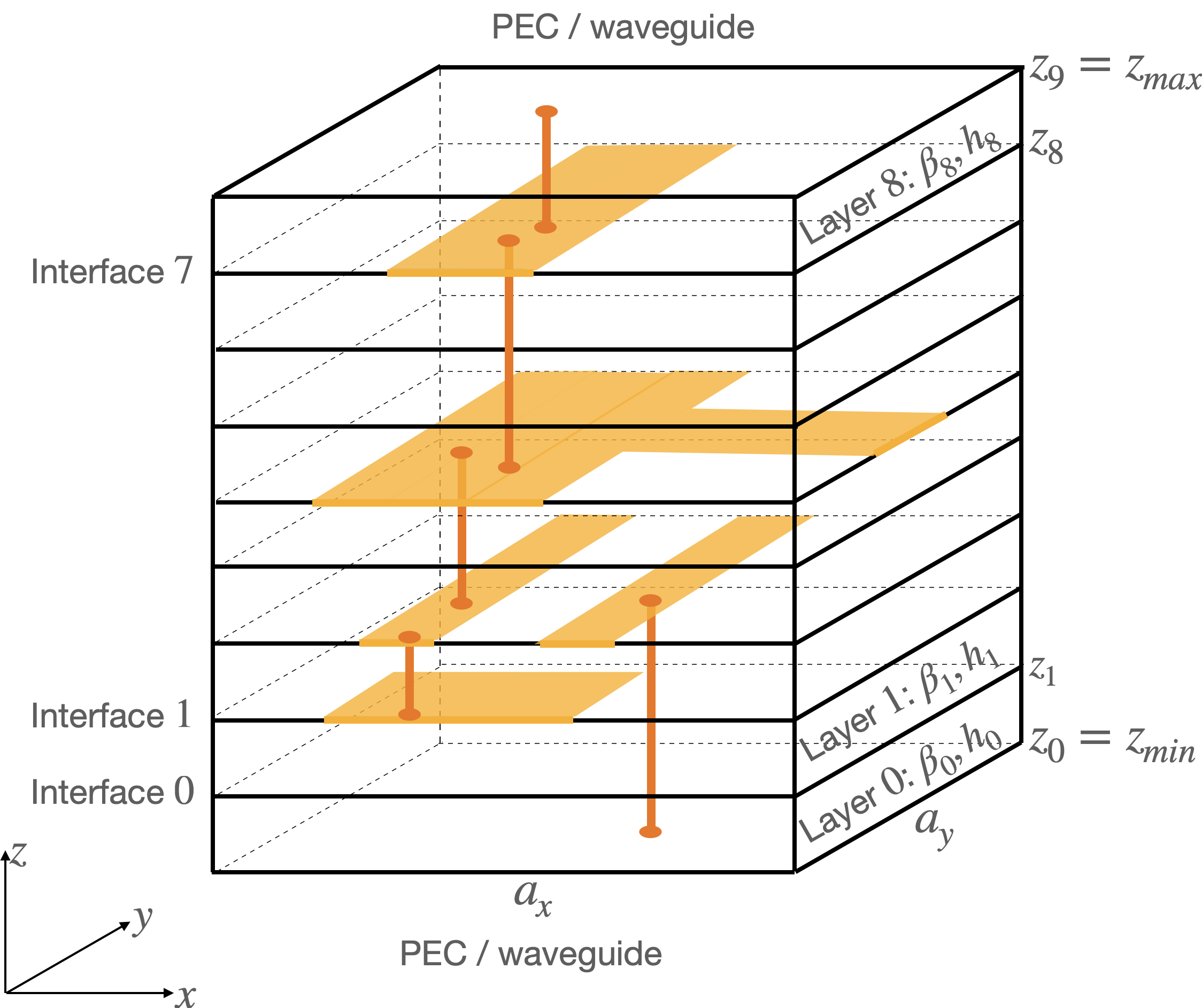}
	\caption{An example PCB structure.}
	\label{fig:Structure}
\end{figure}

Commonly used basis functions include ``rooftop" functions \cite{Sercu1994}, Rao-Wilton-Glisson (RWG) functions \cite{Rao1982}, piecewise sinusoidal functions \cite{Khalil1999}, and full-domain vector functions defined individually for each metallic element (see, e.g., \cite{Melcon1999}). In this study, the proposed methodology is demonstrated using two-dimensional basis functions for planar metallizations \cite{Rautio1987}, as well as volumetric pulse and linear basis functions for vertical interconnects \cite{Rautio2021}. These basis functions are sufficiently simple to allow for the evaluation of overlap integrals as closed-form analytical expressions. This enables the approach described herein to be applied to PCB structures that incorporate both thin planar patch/strip metallizations and vertical interconnect structures, as illustrated in Fig. \ref{fig:Structure}. It should be noted that the expressions for the Green’s function provided below can also be effectively used to derive overlap integrals for other types of basis functions, facilitating the modeling of various structures documented in the literature. Another limitation of the present work stems from the assumption that the longitudinal dimension of the structure is small relative to the wavelengths within the frequency range of interest; consequently, longitudinal currents can be considered essentially uniform along the length of the vertical interconnect elements. Nevertheless, the current approach can be readily extended to account for non-uniform currents such as those occurring in interconnects, vias, and planar metallization layers of finite thickness.

The use of rooftop basis functions for planar metallizations on a uniform grid allows for acceleration via the Fast Fourier Transform (FFT), where the matrix in the projection method is expressed as a sum of block-Toeplitz and block-Hankel matrices. Efficient methods for multiplying such structured matrices by vectors based on transforming them into circulant matrices were developed in \cite{Strang1986, Strang1989}. Consequently, matrix-vector multiplications within this framework can be performed in $O(n \log n)$ operations, requiring the storage of only one row and one column for each type of matrix involved; this results in high overall numerical efficiency when solving for surface and volume currents.

At present, despite the in-depth study of various problems related to calculation of the high-frequency properties of PCBs modeled as shielded multilayer dielectric media containing conductive structures within a rectangular waveguide, there remains a lack of a systematic approach in the open literature for deriving the analytical expressions required for the Galerkin projection method for solving EFIE and for applying it to model arbitrary metal-dielectric PCB structures, including both planar conductors and vertical interconnects. Therefore, the primary objective of this work is to demonstrate a method for calculation of all components of the Green’s function for a shielded layered medium utilizing the scattering matrix formalism \cite{Ko1988, Li1996}, and to explicitly provide the analytical expressions necessary for constructing the matrix used in the Galerkin projection method.

The paper is organized as follows. Section \ref{sec2} presents EFIE formulation of the problem for calculating currents. Section \ref{GF_seq} provides the derivation of the multilayer Green’s function using the S-matrix method. Section \ref{MoM_seq} develops the Galerkin projection method equation, employing surface and volume pulse and linear basis functions, and calculates the overlap integrals for planar metallizations and vertical interconnects. Section \ref{FFT} describes our approach to accelerating computations via the FFT. Finally, Section \ref{Num_ex} presents two numerical examples that demonstrate the capabilities of the method.

\section{Formulation of the problem}\label{sec2}

A general boxed PCB structure is schematically shown in Fig.~\ref{fig:Structure}. A rectangular waveguide with the waveguide axis parallel to Cartesian axis $Z$ and waveguide walls to Cartesian axes $X$ and $Y$ can be filled with an arbitrary multilayer structure with layer interfaces parallel to the $XY$ plane. Metallizations represent planar possibly complex-shaped metal plates, which lie in the vicinity of layer boundaries $z=z_l+0$. The plates can be connected by thin vertical wires (vias), which can cross arbitrary number of interfaces. Electromagnetic properties of such structures are evaluated by solving the boundary-value problem for the time-harmonic Maxwell's equations. In further derivation the factor $\exp(-i\omega t)$ is implicitly assumed, which is a common choice for wave excitation and propagation problems (waveguide modes in our case). The waveguide walls are assumed to be infinitely conducting, while the metallizations and vias may have a finite conductivity, which is taken into account by the Leontovich impedance boundary condition. At the boundaries $z=z_{min}$ and $z=z_{max}$ of the computational domain $\Omega = \{ 0\leq x\leq a_x, 0\leq y\leq a_y, z_{min}<z<z_{max} \}$, one may impose either open-waveguide boundary conditions or conditions corresponding to perfectly conducting end-caps. 

Further we consider the solution of the boundary-value problem by the solution of the integral equation written for the tangential electric field at metallizations and vias \cite{Volakis2012_iem}. This equation follows from the Leontovich boundary conditions \cite{leontovich1948,rytov1940skin}
\begin{equation}
	\Eb_t = Z_{s} \Hb\times\hat{\nb}_s,
\end{equation}
where $Z_{s}$ is the surface impedance, $\hat{\nb}_s$ is the normal to the surface of the conductor. It has the following form:
\begin{equation}\begin{split}
	\hat{\nb}_s\times\left[ \Eb^{exc}(\rb) + i k Z \int_V \Gb(\rb, \rb') \Jb(\rb') dV'\right] = Z_{s} \Jb_s(\rb) 
	\label{EFIE1}
\end{split}\end{equation}
Here $\Gb(\rb, \rb')$ is the dyadic Green's function; $\Jb$ and $\Jb_s$ are the volume and surface electric current densities respectively, $\Eb^{exc}(\rb)$ is the external field that excites the structure at given ports; $k$ and $Z$ are the wave number and impedance at the field observation point $\rb$. Application of the specified boundary condition is justified by the skin depth of the copper conductors being about 2 µm at the frequency of 1 GHz, i.e. approximately an order of magnitude thinner than the thickness of the conductors on printed circuit boards and five orders of magnitude smaller than the vacuum wavelength. It is generally accepted in the works mentioned in the introduction.

Application of MoM to simulation of S-parameters of a PCB consisting of metallizations and vias in a layered waveguide consists of the following steps: projecting the electric current density onto a set of basis functions; projecting the fields onto the same (the Galerking method) or a different set of test functions (Petrov-Galerkin method); calculationg the matrix elements as overlaps between the IE kernel $\Gb(\rb, \rb')$, the basis and the test functions to obtain the MoM linear algebraic system from (\ref{EFIE1}); solving the resulting system of linear algebraic equations for the unknown currents; and calculating the desired output parameters, e.g. S-parameters of the PCB from the solution of the MoM equation. The most resource-consuming steps are the evaluation of the matrix elements and the solution of the linear system. The former must rely on an efficient and robust calculation of $\Gb(\rb, \rb')$ elements, which is the subject of the next section. 

\section{Scattering matrix based computation of fields in a layered waveguide} \label{GF_seq}

The Green's dyadic of a general multilayer structure provides a field at some coordinate $z$ given a `delta'-source at a coordinate $z'$. Instead of deriving its explicit components we derive a procedure of computing fields for given sources; this procedure effectively implements the action of the integral operator in equation (\ref{EFIE1}) and proves to be preferable from a computational standpoint. For this purpose, we use a simple Green's function of a homogeneous waveguide, derive mode amplitudes for a `delta'-source located in a locally homogeneous environment, and apply the scattering matrix method to compute the selfconsistent field at any $z$-level of a multilayer structure.

\subsection{Green's function in a homogeneous waveguide}

Green's function for a homogeneous infinite rectangular waveguide is well known and can be expressed via a summation over the waveguide mode functions \cite{MarcuvitzWH1986}. In a waveguide that is infinite in the longitudinal direction, filled with a homogeneous isotropic medium, and has the cross-section $a_x \times a_y$, the components of the Green's dyadic explicitly read
\begin{equation}
	G^{\alpha \alpha'}(\rb,\rb') = \frac{i}{kZ} \sum_{m,n=0}^{\infty} \mathcal{Z}^{\alpha \alpha'}_{mn} \psi^{\alpha}_{mn} (\rhob) \psi^{\alpha'}_{mn} (\rhob') e^{i \beta_{mn} |z-z'|},
	\label{GF_hom}
\end{equation}
where $\alpha, \alpha' = \{x,y,z\}$. The coefficients in the above equation are
\begin{equation} \begin{split}
	\mathcal{Z}^{\alpha \alpha'} &= \gamma \frac{Z k}{\beta} \left( \frac{k_{\alpha'}^2}{k_{\rho}^2} + \frac{\beta^2}{k^2} \frac{k_{\alpha}^2}{k_{\rho}^2} \right), \\
	\mathcal{Z}^{xy} & = \mathcal{Z}^{yx} = - \gamma \frac{Z k}{\beta} \frac{k_x k_y}{k^2}, \\
	\mathcal{Z}^{zz} &= \gamma \dfrac{Z k}{\beta} \left[ \dfrac{k_{\rho}^2}{k^2} + \dfrac{2i\beta}{k^2} \delta(z-z') \right], \\
	\mathcal{Z}^{z \alpha} &= -\left[ \mathcal{Z}^{\alpha z} = i \mbox{sgn}(z-z') \gamma Z \frac{k_\alpha}{k} \right]
	\label{GF4}
\end{split} \end{equation}
with the superscript $\alpha = \{x,y\}$, and $\alpha,\alpha' = \{x,y\},\{y,x\}$. Here $\rhob = (x,y)$ is a point in the waveguide cross-section, $k_x = \pi m/a_x$, $k_y = \pi n /a_y$ are the transverse wavevector components, $k_{\rho}^2 = k_x^2 + k_y^2$, $\beta = \sqrt{k^2 - k_{\rho}^2}$ is a propagation constant such that the square root branch is taken in the upper half-plane. The constant
\begin{equation}
    \gamma_{mn} = \frac{2s_ms_n}{a_x a_y}
    \label{eq:gamma_const}
\end{equation}
and $s_m = 1/2$ for $m = 0$ and $1$ otherwise.
Here and in what follows, we omit an evident explicit dependence of $k_x$, $k_y$, $k_{\rho}$, $\beta$ and other quantities on the mode indices $m,n$ to shorten formulas. The functions $\psi^{\alpha}_{mn}$ are the waveguide mode functions and explicitly they read:
\begin{equation}\begin{split}
	\psi^x (\rhob) &= \cos(k_x x) \sin(k_y y), \\
	\psi^y (\rhob) &= \sin(k_x x) \cos(k_y y), \\
	\psi^z (\rhob) &= \sin(k_x x) \sin(k_y y). 
\end{split} \end{equation}
These functions participate in the electric field mode expansion, while the magnetic field requires also a fourth function $\chi^z (\rhob) = \cos(k_x x) \cos(k_y y)$.

\subsection{Source emission}

From Eq.~(\ref{GF_hom}) it follows that in a homogeneous waveguide, electric field due to a planar surface current at position $z=z_s$
\begin{equation}
    \Jb(\rb) = \Jb_s(\rhob) \delta (z-z_s)
    \label{eq:Jsurf}
\end{equation}
has the form:
\begin{equation}
\begin{split}
	&\Eb (\rb) = - \frac{i Z}{k} J_{sz}(\rhob) \zhat +\; \\
	&+ \sum_{mn} \sum_{p=e,h} c^{p,\sign(z-z_s)}_{mn} \eb^{p,\sign(z-z_s)}_{mn}(\rhob) e^{i \beta_{mn} |z-z_s|},
\end{split}
\label{eq:emi_field_homo}
\end{equation}
where $p = \{e,h\}$ is polarization index for TM and TE modes respectively,
and the $\pm$ superscript distinguishes modes propagating in the positive and negative direction relative to axis $Z$ for $z > z_s$ and $z < z_s$ respectively. The vector mode wave functions are expressed using the scalar mode functions as \cite{MarcuvitzWH1986}
\begin{equation}\begin{split}
	\eb^{h\pm}(\rhob) &= 
		\begin{pmatrix} 
			- k_{y} \psi^x (\rhob) \\ 
			\;\;\; k_{x} \psi^y (\rhob) \\ 
			0 
		\end{pmatrix}, \\
	\eb^{e\pm} (\rhob) & 
		= \dfrac{Z}{k} \begin{pmatrix} 
			\mp \beta k_{x} \psi^x (\rhob) \\ 
			\mp \beta k_{y} \psi^y (\rhob) \\ 
			\;\;\; i k_{\rho}^2 \psi^z (\rhob) 
			\end{pmatrix},
			\label{e^ehpm_mn} 
\end{split} \end{equation}
and the amplitudes are
\begin{equation} \begin{split}
	c^{e\pm} & = \frac{1}{2} \left( \pm \frac{k_x}{k_{\rho}^2} \mathcal{J}^x \pm \frac{k_y}{k_{\rho}^2} \mathcal{J}^y + \frac{i}{\beta} \mathcal{J}^z \right), \\
	c^{h\pm} & = \frac{1}{2} \frac{k Z}{\beta} \left( \frac{k_y}{k_{\rho}^2} \mathcal{J}^x - \frac{k_x}{k_{\rho}^2} \mathcal{J}^y \right).
	\label{c2}
\end{split} \end{equation}
Here $\mathcal{J}^{\alpha}$ denote the source projections on the scalar mode functions:
\begin{equation} \begin{split}
	\mathcal{J}^{\alpha} &= 2\gamma \intop d \rhob' J^{\alpha} (\rhob') \psi^{\alpha} (\rhob').
	\label{Jalpha}
\end{split} \end{equation}
The expansion of Eq.~(\ref{eq:emi_field_homo}) for an infinite homogeneous waveguide can also be used to find the field in a layered waveguide. In such an inhomogeneous environment the amplitudes $c^{p\pm}_{mn}$ are not self-consistent amplitudes any more, however, they can be used to calculate true self-consistent field amplitudes in conjunction with the scattering matrix method as explained in the next subsections.

\subsection{Scattering matrix method in a waveguide} 

The modal decomposition is a complete expansion of any field in a homogeneous waveguide at some coordinate $z$, provided that there are no sources in some region around this coordinate. The field can be expressed via a discrete set of amplitudes
\begin{equation}
	\Eb (\rb) = \sum_{mn} \sum_{p=h,e} \sum_{s=\pm} a^{p,s}_{mn} \eb^{p,s}_{mn}(\rhob) e^{is\beta_{mn} z}.
	\label{eq:field_expansion}
\end{equation}
Similar series hold for the magnetic field. Eq.~(\ref{eq:field_expansion}) describes a superposition of modes propagating along the $z$ axis. The mode phases in the equation are evaluated relative to the point $z=0$, although one can choose another reference point, which affects the definition of the complex amplitudes $a^{p,s}$. Expansion of Eq.~(\ref{eq:field_expansion}) can be used also for an inhomogeneous waveguide filled with different media with interfaces between different materials parallel to the $xy$ plane, provided that in the vicinity of the expansion point $z$ the medium is locally homogeneous. 

Considering a partition of the waveguide $z_1\leq z\leq z_2$ and denoting the field expansion amplitudes at the bounding coordinates $z_{1,2}$ as $a^{p,s}(z_{1,2})$ one can relate these amplitudes by the S-matrix as follows
\begin{equation}
    \begin{pmatrix} a^{p,-}(z_{1}) \\ a^{p,+}(z_{2})  \end{pmatrix} = S^{p}(z_1,z_2) \begin{pmatrix} a^{p,+}(z_{1}) \\ a^{p,-}(z_{2}) \end{pmatrix}
    \label{eq:smat_def}
\end{equation}
Here and below we assume scattering matrices to be diagonal relative to the mode and polarization indices, thus, to have actual size of $2\times2$. This is possible due to the fact that multilayer background of PCB structures under consideration consists of sections of homogenous dielectric media (possibly lossy), and the modes passing through them do not mix. 

Given two S-matrices of adjacent waveguide partitions $z_1\leq z< z_2$ and $z_2\leq z< z_3$, the S-matrix of the joint partition $z_1\leq z< z_3$ is calculated by the Redheffer star product $\star$ \cite{redheffer1962relation}. Therefore for the purpose of analysis of a boxed PCB with a multilayer background, one only needs to consider the following elementary S-matrices, which are combined by the Redheffer star product:
\begin{enumerate}
	\item S-matrices for $l$-th interface at $z_{l+1}$ between $l$-th and $(l+1)$-th media:
		\begin{equation} \begin{split}
			S^{(I)e}_{l} &= \begin{pmatrix}
				\frac{\beta_l \varepsilon_{l+1} -\beta_{l+1} \varepsilon_l}{\beta_{l+1} \varepsilon_l + \beta_l \varepsilon_{l+1}} & 
				\frac{2 \beta_{l+1} \varepsilon_l}{\beta_{l+1} \varepsilon_l + \beta_l \varepsilon_{l+1}}  \\
				\frac{2 \beta_l \varepsilon_{l+1}}{\beta_{l+1} \varepsilon_l + \beta_l \varepsilon_{l+1}} & 
				\frac{\beta_{l+1} \varepsilon_l -\beta_l \varepsilon_{l+1}}{\beta_{l+1} \varepsilon_l + \beta_l \varepsilon_{l+1}} 
			\end{pmatrix}, \\
			S^{(I)h}_{l} &= \begin{pmatrix}
				\frac{\beta_l \mu_{l+1} -\beta_{l+1} \mu_l}{\beta_{l+1} \mu_l + \beta_l \mu_{l+1}} & 
				\frac{2 \beta_{l+1} \mu_l}{\beta_{l+1} \mu_l + \beta_l \mu_{l+1}}  \\
				\frac{2 \beta_l \mu_{l+1}}{\beta_{l+1} \mu_l + \beta_l \mu_{l+1}} & 
				\frac{\beta_{l+1} \mu_l -\beta_l \mu_{l+1}}{\beta_{l+1} \mu_l + \beta_l \mu_{l+1}} 
			\end{pmatrix};
			\label{interface_smatrices}
		\end{split} \end{equation}
		These matrices relate the mode amplitudes at coordinates $z = z_{l+1} \pm 0$ in Figure 1, so that 
		\begin{equation} \begin{split}
			S^{(I)p}_{l} = S^p (z_{l+1} - 0, z_{l+1} + 0).
		\end{split} \end{equation}
	%
	\item S-matrices for $l$-th homogeneous layer of thickness $h_l$ between the $(l-1)$-th and $l$-th interfaces, not including the interfaces themseves:
		\begin{equation} \begin{split}
			S^{(L)}_{l} = \begin{pmatrix} 0 & e^{i \beta_l h_l} \\ e^{i\beta_l h_l} & 0 \end{pmatrix};
			\label{layer_smatrix}
		\end{split} \end{equation}
		This relation is polarization independent, therefore, a corresponding superscript is omitted. These matrices relate the mode amplitudes at coordinates $z=z_l+0$ and $z_{l+1}-0$ in Figure \ref{fig:Structure}, so that 
		\begin{equation} \begin{split}
			S^{(L)}_{l} = S (z_l + 0, z_{l+1} - 0).
		\end{split} \end{equation}
	\item S-matrices for the top and bottom end-caps (superscript `T' and `B') depending on the boundary conditions at $z=z_{min}$ and $z=z_{max}$:
		\begin{equation} \begin{split}
			S^{(T/B)e/h} &= \begin{pmatrix} 0 & 1 \\ 1 & 0 \end{pmatrix} \mbox{ for an open waveguide};\\
			S^{(T)e/h} &= \begin{pmatrix} \pm 1 & 0 \\ 0 & 0 \end{pmatrix} \mbox{ for a PEC top end-cap};\\
			S^{(B)e/h} &= \begin{pmatrix} 0 & 0 \\ 0 & \pm 1 \end{pmatrix} \mbox{ for a PEC bottom end-cap}.
			\label{topbot_smatrix}
		\end{split} \end{equation}
		The sign "$\pm$" corresponds to the polarization.
\end{enumerate}

\subsection{Self-consistent mode amplitudes in the vicinity of a source} \label{sec:self}

Next let us calculate self-consistent field amplitudes in the vicinity of a planar source located at the coordinate $z=z_s$ given by Eq.~(\ref{eq:Jsurf}). Since the scattering matrices considered here are diagonal relative to the mode and polarization indices, from here on we omit these indices for the S-matrices, mode amplitudes and related Green's function coefficients wherever omitting them does not cause ambiguity in the formulas. 

Suppose that the components of the scattering matrix ${S_{11}(z_s,z_{max})}$ for the waveguide section above the source, and ${S_{22}(z_{min},z_s)}$ for the section below the source have been calculated using elementary S-matrices, and the rules for their multiplication are known. Let us denote the unknown self-consistent mode amplitudes in the vicinity of the source as $a^{\pm}(z_s\pm0)$. Then, these amplitudes can be expressed in terms of the source amplitudes $c^{\pm}(z_s)$, defined in equation (\ref{c2}), as follows:
\begin{equation} \begin{split}
	& \begin{pmatrix}
		a^{-}(z_s-0) \\ a^{+}(z_s+0)
	\end{pmatrix} = 
	\\
	&= D_s \begin{pmatrix}
		{S_{11}(z_s,z_{max})} & 1 \\ 1 & {S_{22}(z_{min},z_s)}
	\end{pmatrix}	
	\begin{pmatrix}
		c^{+}(z_s) \\ c^{-}(z_s)
	\end{pmatrix}
	\label{ac1}
\end{split} \end{equation}
Here the factor
\begin{equation} \begin{split}
	D_s = \frac{1}{1 - {S_{11}(z_s,z_{max})} {S_{22}(z_{min},z_s)}} 
	\label{D1}
\end{split} \end{equation}
appears due to all the multiple reflections from the waveguide sections above and below the source. The remaining two amplitudes are obtained by multiplying by the corresponding reflection coefficients:
\begin{equation} \begin{split}
	a^{-} (z_s+0) &= {S_{11}(z_s,z_{max})} \, a^{+} (z_s+0), \\
	a^{+} (z_s-0) &= {S_{22}(z_{min},z_s)} \, a^{-} (z_s-0).
	\label{self-consistent}
\end{split} \end{equation}
In the latter equations, the amplitude of the self-consistent upward-propagating wave $a^{+}(z_s+0)$ is the sum of the the amplitude of the directly radiated upward-propagating wave from the source $c^{+}(z_s)$ and and the amplitude of a downward-propagating wave $a^{-}(z_s-0)$ reflected from the structure below the source. Similarly, $a^-(z_s - 0)$ can be expressed in terms of $a^+(z_s + 0)$ and $c^-(z_s)$. The amplitude of the self-consistent wave propagating downward above the source, $a^-(z_s + 0)$, is simply the amplitude of the upward-propagating self-consistent wave, $a^+(z_s + 0)$, reflected from the portion of the waveguide above the source.

\subsection{Self-consistent mode amplitudes above and below the source} \label{sec:>}

If the observation point $z_v$ is above the source located at $z_s$, i.e. $z_v > z_s$, then the self-consistent mode amplitudes at the source and observation points are related by an S-matrix of the waveguide section between $z_s$ and $z_v$:
\begin{equation} \begin{split}
	&\begin{pmatrix} a^{-}(z_s+0) \\ a^{+} (z_v) \end{pmatrix} = S(z_s,z_v) \begin{pmatrix} a^{+}(z_s+0) \\ a^{-} (z_v)\end{pmatrix}. 
	\label{ab_ratio>}
\end{split} \end{equation}
If $z_v$ coincides with an interface between different materials, then material where the amplitudes are evaluated should be additionally specified as $z_v \pm 0$.

Using the Redheffer star-product rule for S-matrices, we represent the waveguide section above the source as consisting of two parts: the section between the source and the observation point, and the section above the observation point:
\begin{equation} \begin{split}
	&S_{11}(z_s, z_{max}) = \left[ S(z_s,z_v) \star S(z_v, z_{max}) \right]_{11} = \\
	& = S_{11}(z_s,z_v) + \frac{S_{12}(z_s, z_v) S_{11}(z_v, z_{max}) S_{21}(z_s, z_v)}{1 - S_{22}(z_s, z_v) S_{11}(z_v, z_{max})},
\end{split} \end{equation}
Eq.~(\ref{ab_ratio>}) can be solved for $a^{\pm} (z_v)$ to obtain:
\begin{equation} \begin{split}
	a^{+} (z_v) = & D D^{>} {S_{21}(z_s, z_v)} \left[ c^{+}(z_s) + {S_{22}(z_{min}, z_s)} c^{-}(z_s) \right], \\
	a^{-}(z_v) = & {S_{11}(z_v, z_{max})} a^{+}(z_v).
	\label{b>}
\end{split} \end{equation}
Here $D$ is given in Eq.~(\ref{D1}) and
\begin{equation}
	D^{>} = \frac{1}{1 - S_{22}(z_s, z_v) S_{11}(z_v, z_{max})}.
\end{equation}

Analogously, when the observation point $z_v$ is below the source level $z_s$, i.e. $z_v < z_s$, then
\begin{equation} \begin{split}	
	\begin{pmatrix} a^{-}(z_v) \\ a^{+}(z_s-0) \end{pmatrix} = S(z_v, z_s) \begin{pmatrix} a^{+}(z_v) \\ a^{-}(z_s-0) \end{pmatrix}.
	\label{ab_ratio<}
\end{split} \end{equation}
Solving this equation for $a^{\pm}(z_v)$ using the relation
\begin{equation} \begin{split}
	&S_{22}(z_{min}, z_s) = \left[ S(z_{min},z_v) \star S(z_v, z_s) \right]_{22} = \\
	& = S_{22}(z_v, z_s) + \frac{S_{21}(z_v, z_s) S_{22}(z_{min},z_v) S_{12}(z_v, z_s)}{1 - {S_{22}(z_{min}, z_v)} {S_{11}(z_v, z_s)}}
\end{split} \end{equation}
yields:
\begin{equation} \begin{split}
	a^{-}(z_v) = & DD^{<} {S_{12}(z_v, z_s)} \left( c^{-}(z_s) + {S_{11} (z_s,z_{max})} c^{+}(z_s) \right), \\
	a^{+}(z_v) = & {S_{22}(z_{min}, z_v)} a^{-}(z_v), 
	\label{b<}
\end{split} \end{equation}
where
\begin{equation}
	D^{<} = \frac{1}{1 - {S_{22}(z_{min},z_v)} {S_{11} (z_v, z_s)}}.
\end{equation}

In this subsection we have derived explicit formulas allowing one to relate amplitudes of the emitted and self-consistent field at an arbitrary pair of points within multilayer waveguides. In what follows we use these formulas to provide expressions for the multilayer Green's function components in terms of specific pre-calculated sets of scattering matrices associated with the layers, in which $z_s$ and $z_v$ are located. When the viewing point is the same as the source point we will imply $z_v=z_s+0$.

\subsection{Green's function in a multilayer waveguide}

In a layered medium, we express the observed field at a point $\rb_v$ in the form of a mode superposition together with the 'delta'-source term:
\begin{equation} \begin{split}
	\Eb(\rb_v) &= - \frac{i Z_v}{k_v} J_{z} (\rb_v) \zhat + \\
	&+ \sum_{p=e,h} \sum_{m,n=0}^{\infty} \sum_{\sigma=\pm} a_{mn}^{ps} (z_v) \eb_{mn}^{ps} (\rhob_v).
	\label{field_layered3}
\end{split} \end{equation}

Substituting Eqs.~(\ref{e^ehpm_mn}), (\ref{c2}), (\ref{Jalpha}), (\ref{ac1}) (\ref{self-consistent}), (\ref{b>}), (\ref{b<}) into the above equation, we arrive at an expansion of the Green's function similar to the expansion in a homogeneous waveguide:
\begin{equation} \begin{split}
	- i k_v Z_v & G^{\alpha \alpha'}(\rb_v,\rb_s) = \\
	& = \sum_{m,n=0}^{\infty} \mathcal{Z}^{\alpha \alpha'}_{mn} (z_v,z_s) \psi^{\alpha}_{mn} (\rhob_v) \psi^{\alpha'}_{mn} (\rhob_s),
	\label{GF_layered}
\end{split} \end{equation}
where, suppressing the mode indices and the coordinate dependence,
\begin{equation} \begin{split}	
	\mathcal{Z}^{xx} &= \gamma Z_v \frac{k_v}{\beta_v} \left( C^{e--} \frac{\beta_v^2}{k_v^2} \frac{k_x^2}{k_{\rho}^2} + C^{h++} \frac{\mu_s}{\mu_v} \frac{\beta_v}{\beta_s} \frac{k_y^2}{k_{\rho}^2} \right), \\
	\mathcal{Z}^{yy} &= \gamma Z_v \frac{k_v}{\beta_v} \left( C^{e--} \frac{\beta_v^2}{k_v^2} \frac{k_y^2}{k_{\rho}^2} + C^{h++} \frac{\mu_s}{\mu_v} \frac{\beta_v}{\beta_s} \frac{k_x^2}{k_{\rho}^2} \right), \\
	\mathcal{Z}^{xy} &= \mathcal{Z}^{yx} = \\
	&= \gamma Z_v \frac{k_v}{\beta_v} \left( C^{e--} \frac{\beta_v^2}{k_v^2} - C^{h++} \frac{\mu_s}{\mu_v} \frac{\beta_v}{\beta_s} \right) \frac{k_x k_y}{k_{\rho}^2}, \\
	\mathcal{Z}^{xz \geq} &= \mathcal{Z}^{zx <} = i\gamma Z_v C^{e+-} \frac{\beta_v}{\beta_s} \frac{k_x}{k_v}, \\
	\mathcal{Z}^{yz \geq} &= \mathcal{Z}^{zy <} = \gamma Z_v \frac{k_v}{\beta_s}  C^{e+-} \frac{i k_y \beta_v}{k_v^2}, \\
	\mathcal{Z}^{zx \geq} &= - \mathcal{Z}^{xz <} = \gamma Z_v \frac{k_v}{\beta_v} C^{e-+} \frac{i k_x \beta_v}{k_v^2}, \\
	\mathcal{Z}^{zy \geq} &= - \mathcal{Z}^{yz <} = \gamma Z_v \frac{k_v}{\beta_v} C^{e-+} \frac{i k_y \beta_v}{k_v^2}, \\
	\mathcal{Z}^{zz} &= \gamma Z_v \frac{k_v}{\beta_v} \left[ C^{e++} \frac{\beta_v}{\beta_s} \frac{k_{\rho}^2}{k_v^2} + \frac{2 i \beta_v}{k_v^2} \delta(z_v - z_s) \right].
	\label{mathcalZ_mn} 
\end{split} \end{equation}
The superscripts $\geq$ and $<$ for the $z\alpha$ and $\alpha z$ components, $\alpha = \{ x,y\}$, denote the layouts $z_v \geq z_s$ and $z_v < z_s$. The coefficients $C^{\pm\pm}$ also depend on the relative position of source and observation points, and are given by
\begin{equation} \begin{split}
	C^{p\pm\pm} & = D^p_s D^{p>} {S^p_{21}(z_s,z_v)} \times \\
	&\times \left[ 1 \pm {S^p_{22}(z_{min},z_s)} \right] \left[ 1 \pm {S^p_{11}(z_v, z_{max})} \right]
	\label{C^pmpm>}
\end{split} \end{equation}
for $z_v>z_s$;
\begin{equation} \begin{split}	
	C^{p\pm\pm} & = D^p_{s} D^{p<} {S^p_{12}(z_v,z_s)} \times \\
	&\times \left[ 1 \pm {S^p_{22}(z_{min},z_v)} \right] \left[ 1 \pm {S^p_{11}(z_s,z_{max})} \right]
	\label{C^pmpm<}
\end{split} \end{equation}
for $z_v<z_s$; and
\begin{equation} \begin{split}	
	C^{p\pm\pm} & = D^p_{s} \left[ 1 \pm {S^p_{22}(z_{min},z_v)} \right] \times \\
	&\times \left[ 1 \pm {S^p_{11}(z_v,z_{max})}_{11} \right]
	\label{C^pmpm=}
\end{split} \end{equation}
for $z_v=z_s$. Here the superscript signs `$\pm$' correspond to signs in the first and second parenthesis in the right-hand sides of the latter formulas respectively.

For a pair of $z_v$ and $z_s$, the coefficients $C^{\pm\pm}$ can be conveniently expressed using three sets of aggregated S-matrices associated with layer indices and independent of the exact coordinates $z_s$, $z_v$.

Let $z_v$ be in the layer $l$ and $z_s$ be in the layer $l'$ such that $z_l \leq z_v < z_{l+1}$, and $z_{l'} \leq z_s < z_{l'+1}$. If $z_v$, $z_s$ are at metallizations, then $z_v = z_l$, $z_s = z_{l'}$. Let us define an ``above'' S-matrix as the matrix of the waveguide section between $z_{l+1}$ and $z_{max}$: $S^{A}_{l} = S(z_{l+1}-0,z_{max})$. The section includes the interface $z = z_{l+1}$. Analogously the ``below'' scattering matrix would be $S^{B}_{l} = S(z_{min}, z_l+0)$ and it includes the interface at $z = z_l$. For a pair of layers with indices $l < l'$, a ``pair'' matrix is a scattering matrix of a waveguide section $z_{l+1} \leq z \leq z_{l'}$ including both interfaces $S^{P}_{ll'} = S(z_{l+1}-0,z_{l'}+0)$. If the layers are adjacent, i.e. $l' = l+1$, then $S^{P}_{l,l+1} \equiv S^{(I)}_{l}$.

Now we can express Eqs.~(\ref{C^pmpm>})-(\ref{C^pmpm=}) is terms of the just defined matrices $S^{A}_{l}$, $S^{B}_{l}$ and $S^{P}_{ll'}$ as follows. The common factor $D_s$ responsible for the multiple reflections at the source given by Eq.~(\ref{D1}) becomes
\begin{equation}
	D_{s} = \frac{1}{1 - S^{B}_{l',22} S^{A}_{l',11} e^{2 i \beta_{l'} h_{l'}}}.
	\label{D}
\end{equation}
with $h_{l'}=z_{l'+1}-z_{l'}$ being the layer $l'$ thickness. The other factors vary for different layouts of the source and observation points.

\subsubsection{Observation layer is above the source layer}

$l > l'$, as illustrated in Figure~\ref{fig:S_ABP_of_><}(a). Then, in accordance with the S-matrix composition rule with the star product, one may observe, that
\begin{equation} \begin{split}
	&{S_{11}(z_s,z_v)} = S^{P}_{l'l,11} e^{2 i \beta_{l'} \left(z_{l'+1} - z_s\right)}, \\
	&{S_{21}(z_s,z_v)} = S^{P}_{l'l,21} e^{i \beta_{l'} \left(z_{l'+1} - z_s\right)} e^{i \beta_l (z_v - z_l)}, \\ 
	&{S_{22}(z_s,z_v)} = S^{P}_{l'l,22} e^{2 i \beta_l (z_v - z_l)}, \\
	&{S_{22}(z_{min},z_s)} = S^{B}_{l',22} e^{2 i \beta_{l'} \left(z_s - z_{l'}\right)}, \\
	&{S_{11}(z_v,z_{max})} = S^{A}_{l,11} e^{2 i \beta_l \left(z_{l+1} - z_v\right)}. \\
	\label{C_>1}
\end{split} \end{equation}
and
\begin{equation}
   D^{>} = \frac{1}{1 - S^{A}_{l,11} S^{P}_{l'l,22} e^{2 i \beta_l h_l}}.
   \label{eq:D>_ABP}
\end{equation}
Hence, Eq.~(\ref{C^pmpm>}) transforms to
\begin{equation} \begin{split}
	&C^{\pm\pm} = D_{s} D^{>} S^{P}_{l'l,21} \times \\
	& \times \left[ e^{i \beta_{l'} \left(z_{l'+1} - z_s\right)} \pm S^{B}_{l',22} e^{i \beta_{l'} h_{l'}} e^{i \beta_{l'} \left(z_s - z_{l'}\right)}\right] \times \\
	& \times \left[ e^{i \beta_l (z_v - z_l)} \pm S^{A}_{l,11} e^{i \beta_l h_l} e^{i \beta_l (z_{l+1} - z_v)} \right].
	\label{C_>}
\end{split} \end{equation}

\subsubsection{Observation layer is below source layer}

\begin{figure}[h]
	\centering
	\includegraphics[width=0.45\textwidth]{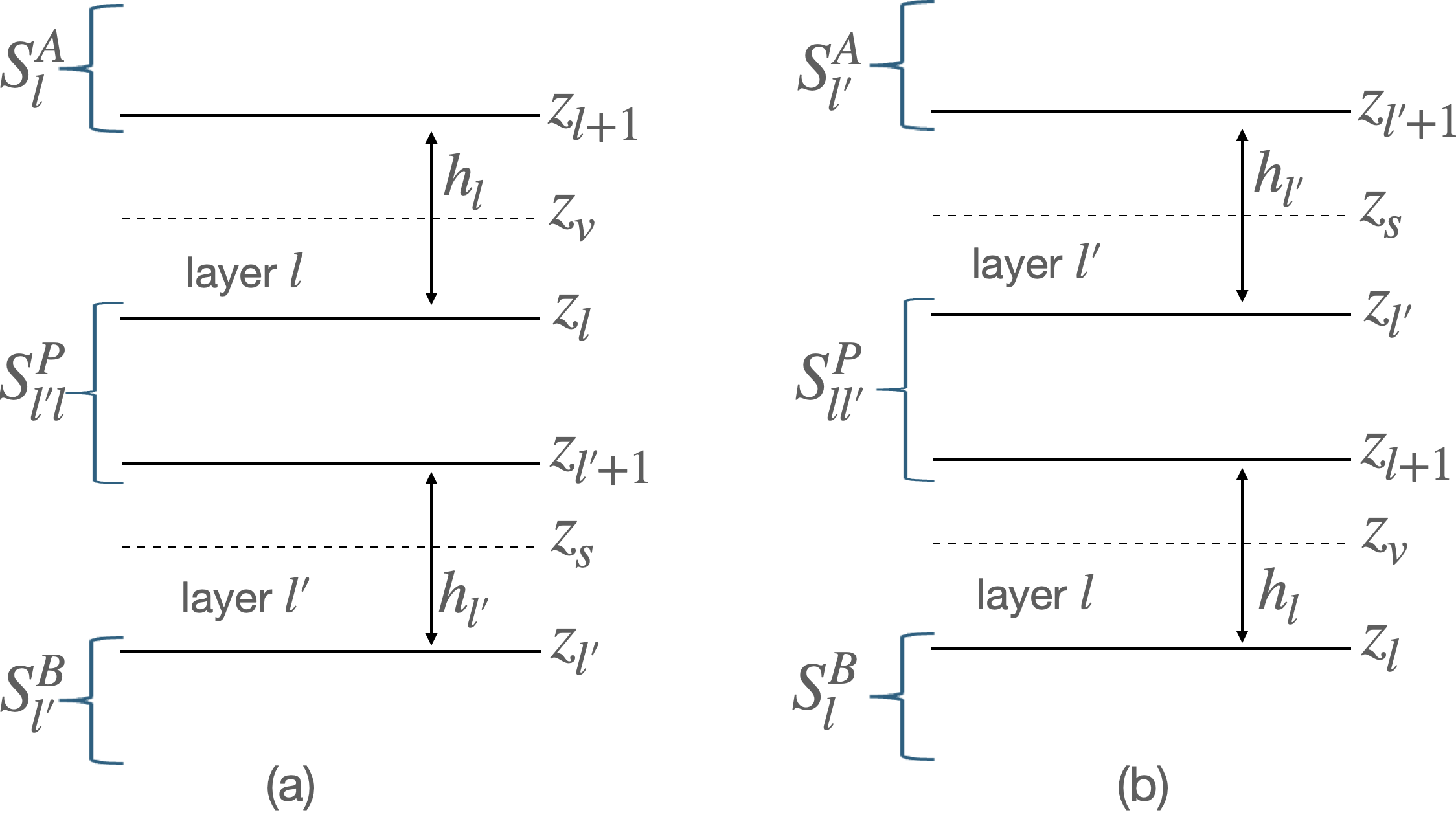}
	\caption{S-matrices associated with layers $l$ and $l'$, when $z_v$ and $z_s$ are in different layers: (a) $z_v > z_s$; (b) $z_v < z_s$.}
	\label{fig:S_ABP_of_><}
\end{figure}

$l < l'$, see Figure~\ref{fig:S_ABP_of_><}(b). In this case
\begin{equation}
    D^{<} = \frac{1}{1 - S^{B}_{l,22} S^{P}_{ll',11} e^{2 i \beta_l h_l}},
    \label{eq:D<_ABP}
\end{equation}
and Eq.~(\ref{C^pmpm<}) rewrites as
\begin{equation} \begin{split}
	&C^{\pm\pm} = D_{s} D^{<} S^{P}_{ll',12} \times \\
	& \times \left[ e^{i \beta_l (z_{l+1} - z_v)} \pm S^{B}_{l,22} e^{i \beta_l h_l} e^{i \beta_l (z_v - z_l)} \right] \\
	& \times \left[ e^{i \beta_{l'} (z_s - z_{l'})} \pm S^{A}_{l',11} e^{i \beta_{l'} h_{l'}} e^{i \beta_{l'} \left(z_{l'+1} - z_s\right)} \right].
	\label{C_<}
\end{split} \end{equation}

\subsubsection{Observation layer and source layer are the same}

\begin{figure}[h]
	\centering
	\includegraphics[width=0.45\textwidth]{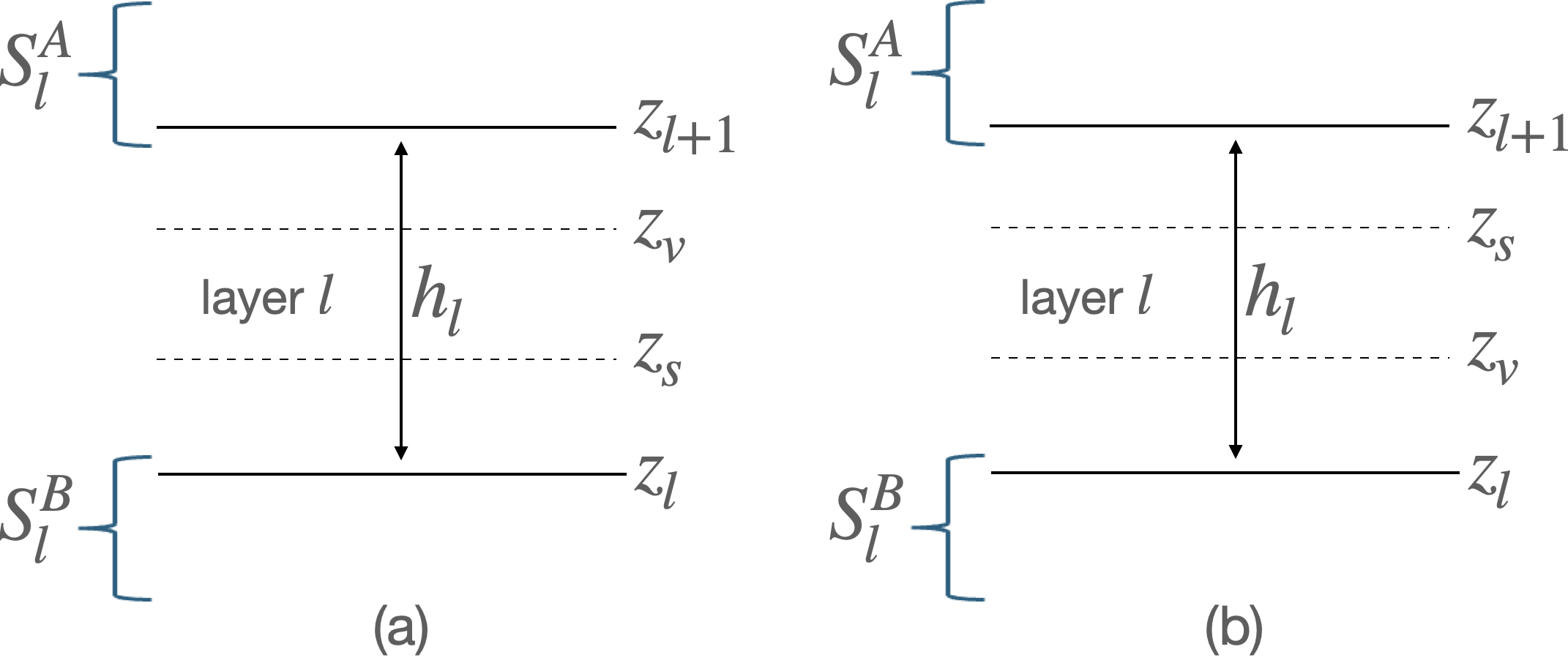}
	\caption{S-matrices associated with layer $l$, when $z_v$ and $z_s$ are in the same layer: (a) $z_v \geq z_s$; (b) $z_v \leq z_s$.}
	\label{fig:S_ABP_of_=}
\end{figure}

$l = l'$, as shown in Figure~\ref{fig:S_ABP_of_=}(a,b). Then, the obvious relations
\begin{equation} \begin{split}
	&D^{>} = D^{<} = 1, \\
	&{S_{21}(z_s,z_v)} = e^{i \beta_l (z_v - z_s)}, \\
	&{S_{12}(z_v,z_s)} = e^{i \beta_l (z_s - z_v)}.
\end{split} \end{equation}
yield
\begin{equation} \begin{split}
	C^{\pm \pm} & = D_{s} \left[ 1 \pm S^{B}_{l,22} e^{2i \beta_l (z_s - z_l)} \right] \times \\
	& \times \left[ 1 \pm S^{A}_{l,11} e^{2i \beta_l (z_{l+1} - z_v)} \right] e^{i \beta_l (z_v - z_s)}
	\label{C=>}
\end{split}\end{equation}
providing that $z_v \geq z_s$, and
\begin{equation} \begin{split}
	C^{\pm \pm} & = D_{s} \left[ 1 \pm S^{B}_{l,22} e^{2i \beta_l (z_v - z_l)} \right] \times \\
	& \times \left[ 1 \pm S^{A}_{l,11} e^{2i \beta_l (z_{l+1} - z_s)} \right] e^{i \beta_l (z_s - z_v)}
	\label{C=<}
\end{split} \end{equation}
when $z_v \leq z_s$. The coefficient for the case when the source and observation coordinates coincide directly follows from a substitution $z_v = z_s$ into any of the latter two equations.

The formulas derived above for the Green's dyadic coefficients have three advantages. First, the exponential factors contain positive layer thinkness constants $h_l$ and variables such as $z_v-z_l$, $z_{l'+1}-z_s$ only. Therefore, all computations including evanescent modes with $\beta_l^2<0$ are stable. Second, the variable source $z_s$ and observation $z_v$ positions appear only in the exponential factors and are not present in $D_s$ and $D^{\lessgtr}$ multipliers, hence, a further integration of such exponents with any polynomial of trigonometric basis functions within the MoM is straightforward. And, third, the coefficients are expressed in a unified way through a fixed set of three types of S-matrices, which can be pre-calculated and stored.

\begin{figure}[h]
	\centering
	\includegraphics[width=0.49\textwidth]{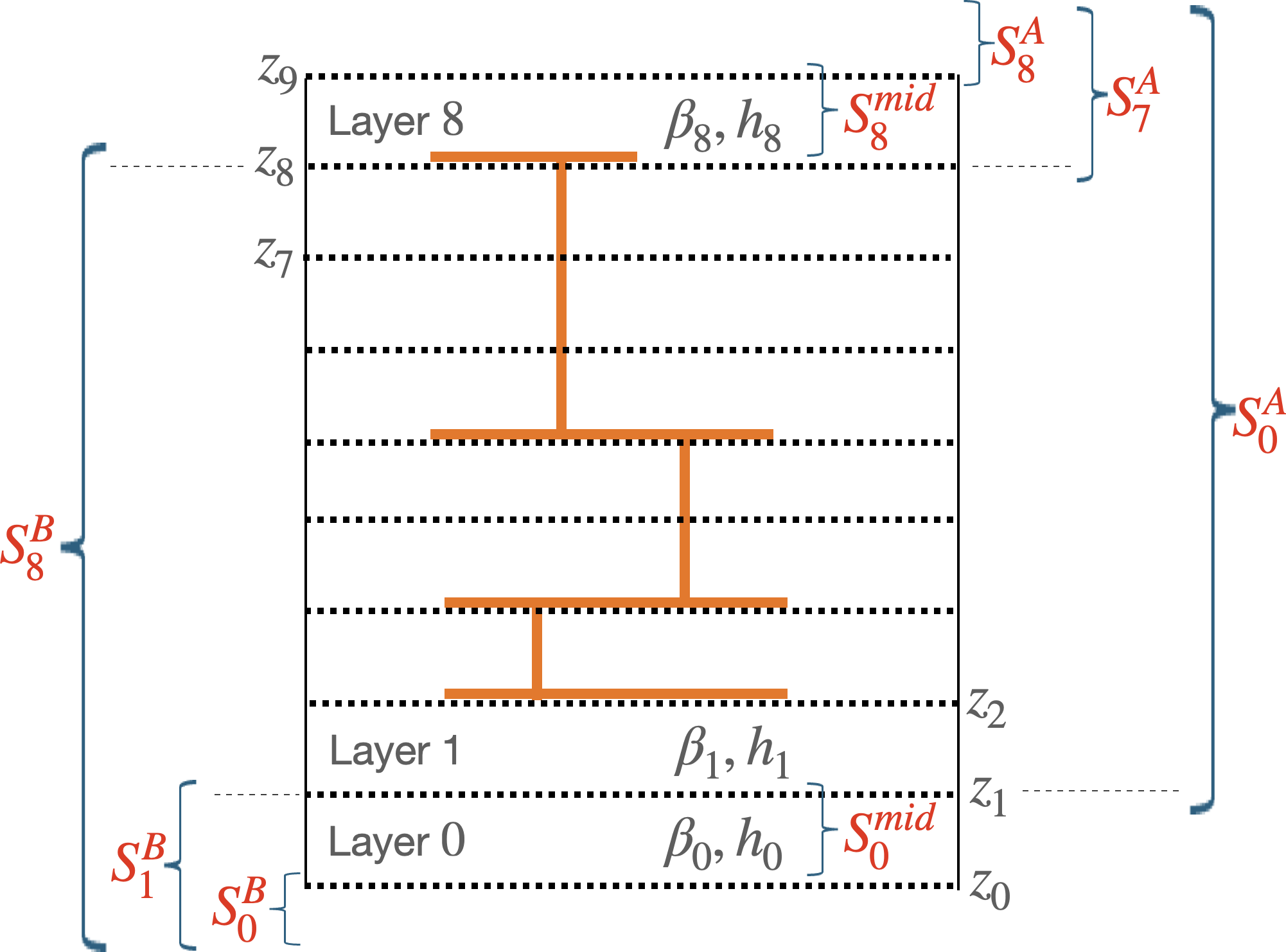}
	\caption{$S^A$ and $S^B$ matrices associated with layers.}
	\label{fig:S_AB_big}
\end{figure}

\begin{figure}[h]
	\centering
	\includegraphics[width=0.45\textwidth]{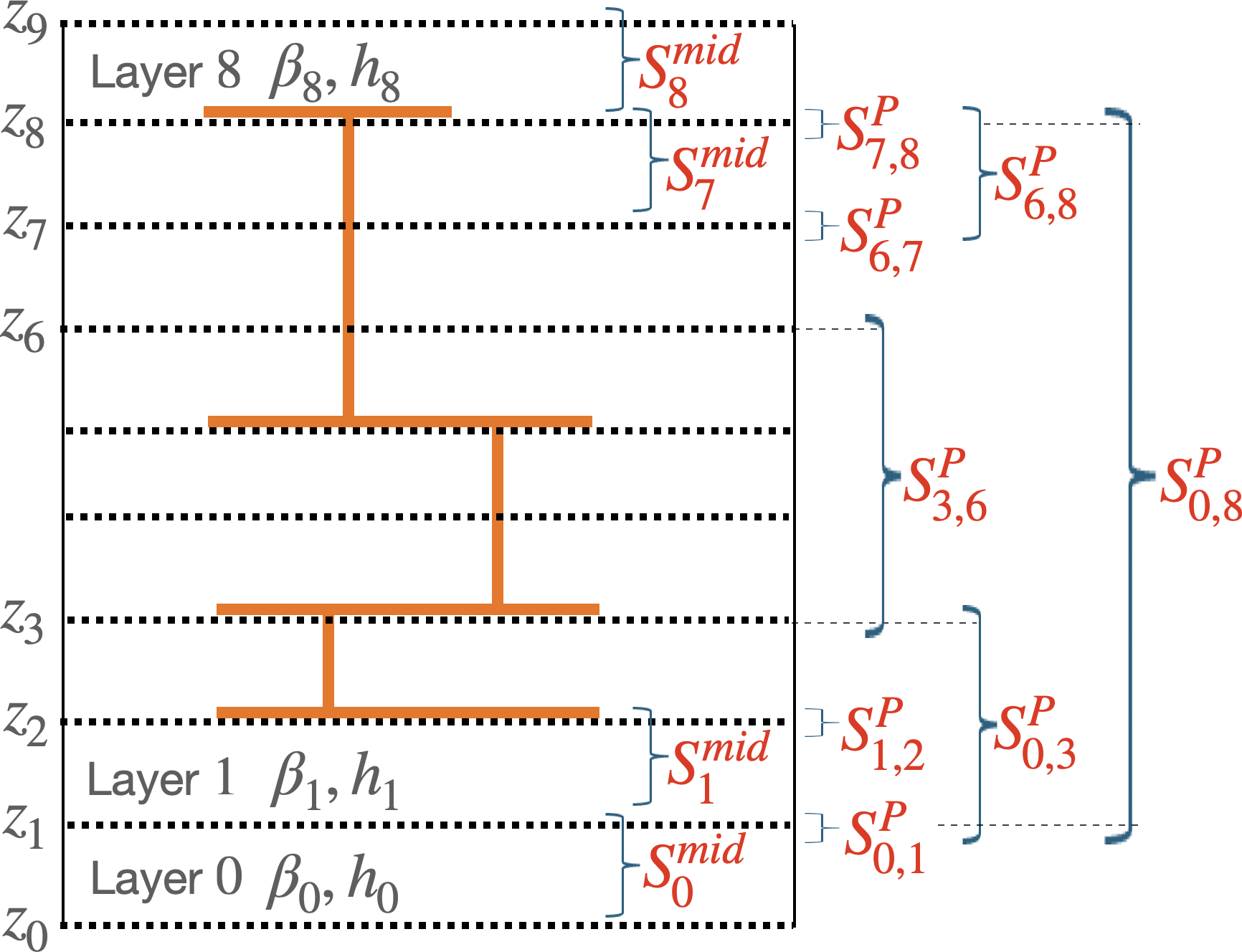}
	\caption{$S^P$ matrices associated with layers.}
	\label{fig:S_P_big}
\end{figure}

To compute the three introduced sets of matrices, $S^A_{l}$, $S^B_l$, and $S^P_{ll'}$ we assume the waveguide to have $(L+1)$ layers, $L$ inner interfaces between different media, and the top and bottom boundary conditions defined in Eq.~(\ref{topbot_smatrix}). As an example, Figures \ref{fig:S_AB_big} and \ref{fig:S_P_big} show a structure with $L=8$ interfaces and 9 layers. For calculations, it is convenient to additionally define matrices $S^{mid}_{l} = S(z_{l}+0,z_{l+1}+0)$, which include the interface at $z = z_{l+1}$ but do not include the interface at $z = z_l$, as illustrated in Figs.~\ref{fig:S_AB_big},\ref{fig:S_P_big}. 
First, in a loop over the layers we calculate the values of $S^{A}_{l,11}$ for $0 \leq l \leq L$. Next, in a second loop we calculate the values of $S^{B}_{l,22}$ for $L \geq l \geq 0$. Finally, in two nested loops we find $S^{P}_{ll''}$ for $0 \leq l \leq (L-1)$, $l+1 \leq l' \leq L$. Matrices $S^{mid}_{l}$ are computed in the first loop and then reused within the following loops.

\section{MoM equations}\label{MoM_seq}

Now we pass to a particular MoM implementation based on the explicit Green's dyadic components derived in the previous section together with the Galerkin method for specific surface and volume basis functions.

\subsection{Basis functions}

In our work, following \cite{Rautio1987,Rautio2020b}, we expand current density on planar metallizations in terms of surface rooftop and half rooftop basis functions on a regular grid. Let us denote the grid size by integers $N_{x}$, $N_{y}$, so that the length elements would be $\Delta x = a_x/N_x$ and $\Delta y = a_y/N_y$. Each full surface rooftop occupies two rectangular mesh elements. Half rooftops are used to model ports and occupy one element of the mesh. The footprint of each interconnect volume basis function on the surface mesh is also one mesh element. The surface current densities and the corresponding expansion coefficients describing the currents in planar PCB traces have physical dimensions of A/m. For the vertical vias we expand the currents in terms of volume pulse and linear basis functions as proposed in \cite{Rautio2020b,Rautio2021}. The volume current densities together with the related expansion coefficients have physical dimensions of A/m$^2$. 

Currents in plane PCB traces located at the coordinate $z_{l+0}$ are expanded in the surface rooftop functions, which are explicitly given by
\begin{equation} 
    \bb^{\alpha}_i (\rb) = b^{\alpha}_i (\rb) \hat{\bm{\alpha}} = t_i (x) p_i (y) \delta(z - z_l) \hat{\bm{\alpha}},
	\label{rooftops} 
\end{equation}
where $\alpha=x,y$. For a vertical via the pulse and linear volume basis functions are:
\begin{equation} \begin{split}
	&\bb^{zP}_j (\rb) = b^{zP}_j (\rb) \zhat = p_j (x) p_j (y) p_j (z) \zhat \\
	&\bb^{zL}_j (\rb) = b^{zL}_i (\rb) \zhat = p_j (x) p_j (y) l_j (z) \zhat.
	\label{volumeBS}
\end{split} \end{equation}
In Eqs.~(\ref{rooftops}) and (\ref{volumeBS}), a pulse function is defined as
\begin{equation} \begin{split}
	& p_i (\alpha) = \left\{ \begin{array}{ll}
		1, & \alpha_i \leq \alpha \leq \alpha_{i+1} \\
		0, & \mbox{otherwise}
	\end{array} \right.;
\end{split} \end{equation}
a full rooftop function is
\begin{equation} \begin{split}
	& t_i (\alpha) = \left\{ \begin{array}{cl}
		\dfrac{\alpha - \alpha_{i-1}}{\alpha_i - \alpha_{i-1}}, & \alpha_{i-1} \leq \alpha \leq \alpha_i \\
		\dfrac{\alpha_{i+1} - \alpha}{\alpha_{i+1} - \alpha_i}, & \alpha_i \leq \alpha \leq \alpha_{i+1} \\
		0, & \mbox{otherwise}
	\end{array} \right..
	\label{surf_roof}
\end{split} \end{equation}
The up-ramp and down-ramp half rooftop functions are essentially linear parts of (\ref{surf_roof}) and are given by either the first or second line in this equation, respectively. The normalized linear function along the vertical coordinate $l_i(z)$ on the interval $[z_i, z_{i+1}]$ has an up-ramp form.
The surface current is expanded as 
\begin{equation} \begin{split}
	\Jb_{surf}(\rb) = \sum_{i=1}^{N^x_b} s^x_i \bb^x_i (\rb) + \sum_{i=1}^{N^y_b} s^y_i \bb^y_i (\rb),
	\label{currents_surf}
\end{split} \end{equation}
and the via current is expanded as
\begin{equation} \begin{split}
	\Jb_{via}(\rb) = \sum_{i=1}^{N^z_b} s^z_i \bb^z_i (\rb),
	\label{currents_via}
\end{split} \end{equation}
where $N^{\alpha}_b$ is the total number of basis functions in the $\alpha$ direction. The total current is
\begin{equation} \begin{split}
	\Jb (\rb_s) = \Jb_{surf}(\rb_s) + \Jb_{via}(\rb_s).
	\label{currents}
\end{split} \end{equation}

\subsection{Overlap integrals}

A MoM linear equation system matrix is derived in a conventional way \cite{Gibson2008}: expansions of Eqs.~(\ref{currents_surf})-(\ref{currents_via}) are substituted into equation (\ref{EFIE1}) and then projected unto the same vector basis functions to yield an algebraic equation system:
\begin{equation} \begin{split}
	\Vb = (\Zb + \Zb_s) \Sb.
	\label{MoM1}
\end{split}\end{equation}
The vector and matrix elements here explicitly write
\begin{equation}\begin{split}
	V^{\alpha}_i = &\intop_V b^{\alpha}_i (\rb_v) E^{\alpha,inc}(\rb_v) dV_v, \label{Galerkin1}
\end{split}\end{equation}
\begin{equation}\begin{split}
	Z^{\alpha \alpha'}_{ik} = &- i k_l Z_l \intop_{V} \intop_{V} b^{\alpha}_i (\rb_v) G^{\alpha \alpha'}(\rb_v, \rb_s) b^{\alpha'}_k (\rb_s) dV_s dV_v, \label{Galerkin2} \\
\end{split}\end{equation}
\begin{equation}\begin{split}
	Z^{\alpha \alpha}_{s,ik} = & Z_{met} \intop_V b^{\alpha}_i (\rb_v) b^{\alpha}_k (\rb_v) dV_v,
	\label{Galerkin3}
\end{split}\end{equation}
and $\Sb$ is the unknown coefficient vector. Similar to the previous section we assume the the observation and source points to be located in layers $l$, $l'$, so that $z_{l'} \leq z_s < z_{l'+1}$ and $z_l \leq z_v < z_{l+1}$. If the source or observation point is on a plane trace, then respectively $z_s = z_{l'}+0$ or $z_v = z_{l}+0$. If the source or observation point are on via, then respectively $z_{l'} < z_s < z_{l'+1}$ or $z_l < z_v < z_{l+1}$.

The components of the Green's function include a part dependent on the source and observation layer indices $l'$ and $l$, which is obtained from the S-matrix method described in the the previous section. The dependence of the Green's function on $\rhob_v$ and $\rhob_s$ is encoded in $\psi^{\alpha}(\rhob_v), \psi^{\alpha'}(\rhob_s)$, while the dependence on $z_v$ and $z_s$ is encoded in the exponential factors of the coefficients  $C^{\pm\pm}$ only. The overlap integrals resulting from the Galerkin procedure where the $\rhob$ and $z$ dependent parts of the Green's function are integrated against the basis functions are evaluated analytically.

The corresponding overlap integrals are denoted as 
\begin{equation} \begin{split}
	Q_{i}^{\alpha} = \intop_V \psi^{\alpha}(\rhob) b^{\alpha}_i (\rb) d \rb, \; \alpha = \{x,y\}
\end{split} \end{equation}
and under the assumption of regular meshes in planes parallel to the $XY$ plane explicitly they are:
\begin{equation} \begin{split}
	Q_{i}^{xF} &= \Delta x \Delta y \sigma^y (\sigma^x)^2 \cos (k_x x_i) \sin (k_y y^c_i), \\
	\label{QF}
\end{split} \end{equation}
for full rooftops;
\begin{equation} \begin{split}
	Q_{i}^{xU} &= \Delta x\Delta y \sigma^y \sin (k_y y^c_i) \times \\ 
		& \times [(\sigma^x)^2 \cos (k_x x_i) / 2 + \eta^x \sin (k_x x_i)], \\
		\label{QU}
\end{split} \end{equation}
for up-ramp half rooftops; and 
\begin{equation} \begin{split}
	Q_{i}^{xD} &= \Delta x \Delta y \sigma^y \sin (k_y y^c_i) \times \\
		& \times [(\sigma^x)^2 \cos (k_x x_i) / 2 - \eta^x \sin (k_x x_i)], \\
		\label{QD}
\end{split} \end{equation}
for down-ramp half rooftops. Here $\alpha^c_i = (\alpha_{i+1} + \alpha_i)/2$ are the cell center coordinates and
\begin{equation} \begin{split}
	& \sigma^{\alpha} = \mathrm{sinc} \left(k_{\alpha} \dfrac{\Delta \alpha}{2} \right), \;
	\eta^{\alpha} = \dfrac{k_{\alpha} \Delta \alpha - \sin (k_{\alpha} \Delta \alpha)}{ (k_{\alpha}\Delta  \alpha)^2}.
\end{split} \end{equation}
The latter two functions are convenient both from the analytical and computational viewpoints. Expressions for $Q_{i}^{yF}$, $Q_{i}^{yU}$, $Q_{i}^{yD}$ are analogous are are obtained by switching the $x$ and $y$ coordinates.

For volumetric via basis functions, the in-plane part of the overlap integrals is 
\begin{equation} \begin{split}
	&Q^{\rho}_{i} = \intop_S \psi^z(\rhob) p_i (x) p_i (y) d\rhob =\\
	& =  \Delta x \Delta y \; \sigma^x \sigma^y \sin(k_v x_i^c) \sin(k_y y_i^c).
	\label{Qxy}
\end{split} \end{equation}
The vertical part, i.e. integrals over $z_v$ and $z_s$ of the exponential factors in coefficients $C^{\pm\pm}$ with functions $p_i (z)$ and $l_i (z)$, are concisely expressed using the following auxiliary functions:
\begin{equation} \begin{split}
	\zeta_P(u) & = i (1 - e^{i u}) / u, \\
	\zeta^{+}_L (u) &= [(1 - i u) e^{i u} - 1] / u^2, \\
	\zeta^{-}_L (u) &=  (1 + i u - e^{i u}) / u^2.
\end{split} \end{equation}
With these notations, the single $z$-overlap integrals are expressed as:
\begin{equation} \begin{split}
	Q^{zP\pm}_{i} &= \intop_{z_l}^{z_{l+1}} p_i (z) e^{i \beta_l \Delta z_{\pm}} dz = h_l \zeta_P (\beta_l h_l),\\
	Q^{zL\pm}_{i} &= \intop_{z_l}^{z_{l+1}} l_i (z) e^{i \beta_l \Delta z_{\pm}} dz = h_j \zeta_L^{\pm} (\beta_l h_l),\\
	\label{IzPLi}
\end{split} \end{equation}
where $\Delta z_+=z-z_l$, $\Delta z_-=z_{l+1}-z$.
In addition, for source and observation point both on a via in the same layer, the following double overlap integrals need to be evaluated:
\begin{equation} \begin{split}
	&Q^{zzPP+-}_{ik} 
	= \intop_{z_l}^{z_{l+1}} \intop_{z_l}^{z_{l+1}} p_i (z) p_k (z') e^{i \beta_l |z-z'|} dz' dz = \\
	&\hspace{2in} = 2 h_l^2 \zeta^-_L (\beta_l h_l), \\
	&Q^{zzPP-+}_{ik} = \intop_{z_l}^{z_{l+1}} \intop_{z_l}^{z_{l+1}} p_i (z) p_k (z') e^{i \beta_l (h_l - |z-z'|)} dz' dz = \\
	&\hspace{2in} = 2 h_l^2 \zeta^+_L ( \beta_l h_l), \\
\end{split} \end{equation}
\begin{equation} \begin{split}
	&Q^{zzLP+-}_{ik} = Q^{zzPL+-}_{ik} = \\
	&\hspace{0.2in} = \intop_{z_l}^{z_{l+1}} \intop_{z_l}^{z_{l+1}} l_i (z) p_k (z') e^{i \beta_l |z-z'|} dz' dz = \\
	&\hspace{2in} = h_l^2 \zeta^-_L (\beta_l h_l), \\
	&Q^{zzLP-+}_{ik} = Q^{zzPL-+}_{ik} = \\
	&\hspace{0.2in} = \intop_{z_l}^{z_{l+1}} \intop_{z_l}^{z_{l+1}} l_i (z) p_k (z') e^{i \beta_l (h_l - |z-z'|)} dz' dz = \\
	&\hspace{2in} = h_l^2 \zeta_L^+ (\beta_l h_l), \\
\end{split} \end{equation}
\begin{equation} \begin{split}
	&Q^{zzLL+-}_{ik} = \intop_{z_l}^{z_{l+1}} \intop_{z_l}^{z_{l+1}} l_i (z) l_k (z') e^{i \beta_l |z-z'|} dz' dz = \\
	&\hspace{0.5in} = 2 h_l^2 \frac{\left( \frac{1}{2} + \frac{1}{3} i \beta_l h_l \right) - \zeta_{L}^+ (\beta_l h_l)}{(\beta_l h_l)^2}, \\
	&Q^{zzLL-+}_{ik} = \intop_{z_l}^{z_{l+1}} \intop_{z_l}^{z_{l+1}} l_i (z) l_k (z') e^{i \beta_l (h_l - |z-z'|)} dz' dz = \\
	&\hspace{0.5in} = 2 h_l^2 \frac{\left( \frac{1}{2} - \frac{1}{3} i \beta_l h_l \right) e^{i \beta_l h_l} - \zeta_{L}^- (\beta_l h_l)} {(\beta_l h_l)^2},
\end{split} \end{equation}	
\begin{equation} \begin{split}	
	Q^{zzPP, \delta}_{ik} &= \intop_{z_l}^{z_{l+1}} \intop_{z_l}^{z_{l+1}} p_i (z) p_k (z') \delta(z-z') dz' dz = \frac{1}{h_l}, \\
	Q^{zzLP, \delta}_{ik} &= Q^{zzPL, \delta}_{ik} = \\
	& =\intop_{z_l}^{z_{l+1}} \intop_{z_l}^{z_{l+1}} l_i(z) p_k(z') \delta(z-z') dz' dz = \frac{1}{2 h_l}, \\
	Q^{zzLL, \delta}_{ik} &= \intop_{z_l}^{z_{l+1}} \intop_{z_l}^{z_{l+1}} l_i(z) l_k(z') \delta(z-z') dz' dz = \frac{1}{3 h_l}.
\end{split} \end{equation}	

\subsection{MoM matrix}

Elements $Z^{\alpha \alpha'}_{ik}$ of the matrix $\Zb$ are calculated using equation (\ref{Galerkin2}). For currents and observation points both located at plane PCB traces the matrix elemets explicitly read
\begin{equation}\begin{split}
	Z^{\alpha \alpha'(\cdot)(\star)}_{ik} = \sum_{mn} \mathcal{Z}^{\alpha \alpha'}_{mn} Q_{i,mn}^{\alpha (\cdot)} Q_{k,mn}^{\alpha'(\star)},\;\alpha, \alpha' = \{x,y\},
	\label{Z^alpha_alpha'_ik}
\end{split} \end{equation}
where $(\cdot), (\star) = \{ U, D, F \}$ for up-ramp, down-ramp and full rooftops, and $\mathcal{Z}^{\alpha \alpha'}_{mn}$ follow from (\ref{mathcalZ_mn}), (\ref{C_>}) and (\ref{C_<}) by associating $z_v = z_l$, and $z_s = z_{l'}$.

For currents at vertical vias and observation points at plane PCB traces one gets
\begin{equation} \begin{split}
	Z^{\alpha(\star) z(\cdot)}_{ik} 
	&=  \sum_{nm} Q^{\alpha(\star)}_{i,mn} Q^{\rho}_{k,mn} \times \\
	& \times \left( \mathcal{Z}^{\alpha z +}_{mn} Q^{z(\cdot)+}_{k,mn} + \mathcal{Z}^{\alpha z -}_{mn} Q^{z(\cdot)-}_{k,mn} \right);
	\label{Z^alphaz_ik}
\end{split} \end{equation}		
and for currents at plane PCB traces with observation points at vertical vias:
\begin{equation} \begin{split}
	Z^{z(\cdot) \alpha(\star)}_{ik}  
		&= \sum_{nm} Q^{\rho}_{i,mn} Q^{\alpha(\star)}_{k,mn} \times \\
		& \times \left( \mathcal{Z}^{z\alpha+}_{mn} Q^{z(\cdot)+}_{i,mn} + \mathcal{Z}^{z\alpha-}_{mn} Q^{z(\cdot)-}_{i,mn} \right).
		\label{Z^zalpha_ik}
\end{split} \end{equation}	
In the latter two equations $\alpha = \{x,y\}$, and $(\star) = \{ U, D, F \}$, while $(\cdot) = \{ P, L \}$ for via pulse and linear functions. $\mathcal{Z}^{\alpha z \pm}$ and $\mathcal{Z}^{z \alpha \pm}$ follow from (\ref{mathcalZ_mn}), (\ref{C_>}), and (\ref{C_<}) by taking $z_v = z_l$ and $z_s = z_{l'}$ for the observation and source points.

When the via current and observation points at via segments are located in different layers on obtains
\begin{equation} \begin{split}
	&Z^{zz(\cdot)(\star)}_{ik} 
		= \sum_{nm} Q^{\rho}_{i,mn} Q^{\rho}_{k,mn} \times \\
		& \times \left( \mathcal{Z}^{zz+-}_{mn} Q^{z(\cdot)+}_{i,mn} Q^{z(\star)-}_{k,mn}
		+ \mathcal{Z}^{zz++}_{mn} Q^{z(\cdot)+}_{i,mn} Q^{z(\star)+}_{k,mn} + \right. \\
		& \left. \;\;\; + \mathcal{Z}^{zz--}_{mn} Q^{z(\cdot)-}_{i,mn} Q^{z(\star)-}_{k,mn}
		+ \mathcal{Z}^{zz-+}_{mn} Q^{z(\cdot)-}_{i,mn} Q^{z(\star)+}_{k,mn} \right),
	\label{Z^zz_ik}
\end{split} \end{equation} 
with $(\cdot), (\star) = \{ P, L \}$. Finally, if the current and observation points are on vertical via segments located in the same layer the matrix coefficient is
\begin{equation} \begin{split}
	&Z^{zz(\cdot)(\star)}_{ik} = \sum_{mn} Q^{\rho}_{i,mn} Q^{\rho}_{k,mn} \left[ \frac{2 i Z_v}{k_v} Q^{zz(\cdot)(\star), \delta}_{ik,mn} + \right. \\
	& \hspace{0.2in} \left. + \mathcal{Z}^{zz+-}_{mn} Q^{zz(\cdot)(\star)+-}_{ik,mn} 
	+ \mathcal{Z}^{zz++}_{mn} Q^{z(\cdot)+}_{i,mn} Q^{z(\star)+}_{k,mn} + \right. \\
	& \hspace{0.2in} \left. + \mathcal{Z}^{zz--}_{mn} Q^{z(\cdot)-}_{i,mn} Q^{z(\star)-}_{k,mn}   
	+ \mathcal{Z}^{zz-+}_{mn} Q^{zz(\cdot)(\star)-+}_{ik,mn} \right],
	\label{ZzzPPLPLL=}
\end{split} \end{equation}
where $(\cdot), (\star) = \{ P, L \}$.

\subsection{Vias crossing several layers}

In the equations of the previous section we have implicitly considered vias and corresponding basis functions occupying a single layer. In general a vertical interconnect can cross several waveguide layers. In this case, additional boundary conditions need to be imposed for the currents to be continuous at the junctions of the interconnect basis functions at the layer boundaries. In our work we assume that the length of any via is small enough so that the current flowing through it can be assumed constant. In this case, for each via, we can define only two basis functions, pulse and linear, occupying the whole via length similarly to the previous section.

Suppose that a $i$-th via crosses $n_i$ layers from $z_{l_i}$ to $z_{l_i + n_i}$, see Figures \ref{fig:S_AB_big} and \ref{fig:S_P_big}. Then the whole via can be projected onto two volume basis functions that span $n_i$ layers:
\begin{equation} \begin{split}
	b^{zP}_{i} & = p_i (x) p_i(y) \sum_{l=l_i}^{l_i+q_i-1} p_l (z), \\
	b^{zL}_{i} & = p_i (x) p_i(y) \sum_{l=l_i}^{l_i+q_i-1} \left[ \frac{z_l - z_{l_i}}{h^{via}_i} p_l (z) + \frac{h_l}{h^{via}_i} l_l (z) \right].
	\label{aggregate_vias}
\end{split} \end{equation}
where $h^{via}_i = z_{l_i + q_i} - z_{l_i}$ is the total length of $i$-th via.

Summations in equations (\ref{aggregate_vias}) lead to weighted single summations for the $Z^{z \alpha}$ and $Z^{\alpha z}$ elements and weighted double summations for the $Z^{zz}$ elements in the MoM matrix.

\subsection{Additional terms due to finite conductivity}

The second matrix in the MoM Eq.~(\ref{MoM1}) is non-zero in the case of a finite surface conductivity of plane PCB traces, and its entries are defined by Eq.~(\ref{Galerkin3}). These entries are non-zero for $xx$, $yy$ and $zz$ components only. For planar metallization layers, non-zero contributions arise from the overlap of planar basis functions with one another:
\begin{equation} \begin{split}
	Z^{\alpha\alpha UU}_{s,i_{\alpha} i_{\alpha}} &= Z^{\alpha\alpha,DD}_{s,i_{\alpha} i_{\alpha}} = \frac{1}{3}Z_{met}\Delta x \Delta y, \\
	Z^{\alpha\alpha FF}_{s,ii} &= \frac{2}{3} Z_{met} \Delta x \Delta y, \\
	Z^{\alpha\alpha UD}_{s,i_{\alpha},i_{\alpha}-1} &= Z^{\alpha\alpha,FF}_{s,i_{\alpha},i_{\alpha}-1} = \\
	= Z^{\alpha\alpha DU}_{s,i_{\alpha},i_{\alpha}+1} &= Z^{\alpha\alpha,FF}_{s,i_{\alpha},i_{\alpha}+1} = \frac{1}{6} Z_{met} \Delta x \Delta y.
\end{split} \end{equation}
The surface impedance $Z_{met}$ can be estimated from a volume conductivity \cite{Rautio2020}. The indices $i_{\alpha}$ and $i_{\alpha}+1$ denote the neighbouring $\alpha$-directed mesh elements for overlapping up- and down-ramped half rooftops and overlapping halves of the nearby full rooftops.

For $zz$ components, the only non-zero contributions come from the overlap of the via basis functions defined on the same elements:
\begin{equation} \begin{split}
	Z^{zz PP}_{s,ii} &= Z_{met} \Delta x \Delta y h^{via}_i, \\
	Z^{zz PL}_{s,ii} &= Z^{zz,LP}_{s,ii} = \frac{1}{2} Z_{met} \Delta x \Delta y h^{via}_i, \\
	Z^{zz LL}_{s,ii} &= \frac{1}{3} Z_{met} \Delta x \Delta y h^{via}_i.
\end{split} \end{equation}

\subsection{Units and scaling}

Inspecting physical dimensions of the matrix and vector elements obtained in the previous three sections, we note that in equations (\ref{MoM1})-(\ref{Galerkin3}) the MoM matrix entries $Z^{\alpha\alpha'}_{ik}$ and $Z^{\sigma,\alpha\alpha'}_{ik}$ have the units of $\Omega$m$^2$ for $\alpha, \alpha' = \{xx, yy, xy, yx\}$ and $\Omega$m$^3$ for $\alpha, \alpha' = \{xz, yz, zx, zy\}$. Entries $Z^{zz}_{ik}$, $Z^{\sigma,zz}_{ik}$ have the units of $\Omega$m$^4$. The source terms $V^{\alpha}_i$ have units of Vm for $\alpha = \{x,y\}$ and Vm$^2$ for $\alpha = z$. Finally, as previously mentioned, $s^{\alpha}_k$ have dimensions of A/m for $\alpha = \{x,y\}$ and A/m$^2$ for $\alpha = z$.

Using fixed dimensions for the grid elements, one can introduce the following normalization of the terms in the equation (\ref{MoM1}):
\begin{equation} \begin{split}
	&\mathcal{Z}^{\alpha \alpha'}_{ik} \mapsto \frac{a_x a_y}{(\Delta x \Delta y)^2} \mathcal{Z}^{\alpha \alpha'}_{ik}, \;\;\; \alpha, \alpha' = \{x,y\}, \\	
	&\mathcal{Z}^{\alpha z}_{ik} \mapsto \frac{a_x a_y}{(\Delta x \Delta y)^2 h^{via}_{k}} \mathcal{Z}^{\alpha z}_{ik}, \\
	&\mathcal{Z}^{z \alpha}_{ik} \mapsto \frac{a_x a_y}{(\Delta x \Delta y)^2 h^{via}_{i}} \mathcal{Z}^{z \alpha}_{ik}, \\
	&\mathcal{Z}^{zz}_{ik} \mapsto \frac{a_x a_y}{(\Delta x \Delta y)^2 h^{via}_{i} h^{via}_{k}} \mathcal{Z}^{zz}_{ik}, \\
	&Z^{\sigma, xx}_{ik} \mapsto \frac{1}{\Delta y^2} Z^{\sigma, xx}_{ik}, \; 
	Z^{\sigma, yy}_{ik} \mapsto \frac{1}{\Delta x^2} Z^{\sigma, yy}_{ik}, \\
	&Z^{\sigma, zz}_{ik} \mapsto \frac{1}{(\Delta x \Delta y)^2} Z^{\sigma, zz}_{ij}, \\
	&V^x_i \mapsto \Delta y V^x_i, \;
	V^y_i \mapsto \Delta x V^y_i, \;
	V^z_i \mapsto  \Delta x \Delta y V^z_i, \\
	&s^x_k = \frac{1}{\Delta y} I^{x}_k, \;
	s^y_k = \frac{1}{\Delta x} I^{y}_k, \;
	s^z_k = \frac{1}{\Delta x \Delta y} I^z_k.
\end{split} \end{equation}

Then equation (\ref{MoM1}) transforms to:
\begin{equation} \begin{split}
	\Vb = \Zb_{tot} \Ib,
	\label{MoM2}
\end{split}\end{equation}
where elements of $\Zb_{tot} = \Zb + \Zb_s$, $\Vb$ and $\Ib$ are measured respectively in ohms, volts and amperes.

\section{FFT accelleration} \label{FFT}

It is well known that the infinite series over the waveguide modes in the $\Zb_{tot}$ matrix elements are slowly converging, especially for $z_v = z_s$, requiring summation of thousands of modes for convergence. Also, the resulting MoM matrix can be large for complex shielded PCB problems. The matrix is dense, which in practice prohibits the application of direct inversion methods and even application of iterative linear system solvers, such as BiCG or GMRES, is prohibitively expensive. On an equidistant grid, both the matrix filling and solving procedures can be accelerated using the FFT-based techniques. These techniques are described, e.g. in \cite{Rautio2014a,Rautio2015,Okhmatovski2024}. In this study, we also consider this idea, however, with a slight modification of the previously proposed scheme. We do not subdivide each rectangular element into four subelements as other authors did. 
 
Since the full expressions for all matrix elements cast into a form suitable for the application of the Fast Fourier Transform (FFT) are quite cumbersome, the derivation is presented here using the matrix element $Z^{xx}_{ik}$ for a pair of full rooftop basis functions as an example. The derivation of analogous expansions for half rooftop basis functions, as well as for the elements $Z^{yy}_{ik}$, $Z^{xy}_{ik}$, $Z^{\alpha z}_{ik}$, and $Z^{z \alpha}_{ik}$, follows the same logic. FFT-based acceleration is not applied to the elements $Z^{zz}_{ik}$, as the computational cost of calculating them is significantly lower than for the other elements.

On a regular mesh, if we denote the node indices of the mesh in the $\alpha = \{x,y\}$ directions by $i_{\alpha} = 0,\dots, N_{\alpha}-1$, where $N_{\alpha}$ is the number of mesh elements in the $\alpha$ direction, then
\begin{equation}
	\alpha_{i_{\alpha}} = i_{\alpha} \Delta \alpha, \thinspace \alpha^c_{i_{\alpha}} = \left( i_{\alpha} + \dfrac{1}{2} \right)\Delta \alpha,
\end{equation}
and
\begin{equation}
	k_{\alpha} \alpha_{i_\alpha} = 2 \pi \dfrac{k_{\alpha} m}{2N_{\alpha}}.
\end{equation}
Denoting the complex root of unity as
\begin{equation}
	\omega_{N}^{n}=e^{2\pi in/N},
\end{equation}
and using it's well-known properties
\begin{equation} \begin{split}
	&\omega_{kN}^{kn}=\omega_{N}^{n}, \\
	&\omega_{N}^{n_{1}}\omega_{N}^{n_{2}}=\omega_{N}^{n_{1}+n_{2}},
\end{split} \end{equation}
we can expand the overlap integral $Q_{i,mn}^{xF}$ as follows 
\begin{equation} \begin{split}
	Q_{i,mn}^{xF} &= \dfrac{1}{4i}\Delta x\Delta y \sigma_n^y  (\sigma_m^x)^2\left(\omega_{2 N_x}^{i_x m}+\omega_{2 N_x}^{-i_x m}\right) \times \\
	& \times \left(\omega_{2 N_y}^{\left(i_y +1/2\right)n}-\omega_{2 N_y}^{-\left(i_y +1/2\right)n}\right).
\end{split} \end{equation}
Introducing additional integer variable $\zeta_{\alpha}\geq 1$ to control the Fourier transform accuracy, subdividing the Fourier summations over $m$ and $n$ into double summations over $m,m'$ and $n,n'$, and rearranging them one can obtain
\begin{equation} \begin{split}
	&Z^{xxFF}_{ik} = \sum_{mn} \mathcal{Z}_{mn}^{xx} Q_{i,mn}^{xF} Q_{k,mn}^{xF} 
		= - \frac{1}{16} (\Delta x \Delta y)^2 \times \\
		& \times \sum_{n=0}^{2 N_y - 1} \sum_{m = 0}^{2 N_x-1} \left( \tilde{E}_{mn} \omega_{2 N_x}^{-(i_x + k_x) m} \omega_{2 N_y}^{-(i_y + k_y + 1) n} \right. - \\ 
		& \left. \hspace{0.8in} - \tilde{E}_{mn} \omega_{2 N_x}^{-(i_x + k_x) m} \omega_{2 N_y}^{-(i_y - k_y) n} \right. + \\
		& \left. \hspace{0.8in} + \tilde{E}_{mn} \omega_{2 N_x}^{-(i_x - k_x) m} \omega_{2 N_y}^{-(i_y + k_y + 1) n} \right. - \\
		& \left. \hspace{0.8in} - \tilde{E}_{mn} \omega_{2 N_x}^{-(i_x - k_x) m} \omega_{2 N_y}^{-(i_y - k_y) n} \right),
		\label{Z^xxFF_ik_FFT}						
\end{split} \end{equation}	
where
\begin{equation} \begin{split}
	\tilde{E}_{00} &= 4 \sum_{n'=0}^{\zeta_y-1} \sum_{m' = 0}^{\zeta_x-1} \mathcal{Z}_{2 N_x m', 2N_y n'}^{xx} (\sigma_{2N_y n'}^y)^2 (\sigma_{2 N_x m'}^x)^4, 
\end{split} \end{equation}	
for $m \neq 0$
\begin{equation} \begin{split}
	\tilde{E}_{m0} &= \sum_{n'=0}^{\zeta_y-1} \sum_{m' = 0}^{\zeta_x-1} 2 (\sigma_{2N_y n'}^y)^2 \times \\
	& \times \left[ \mathcal{Z}_{2 N_x m' + m, 2N_y n'}^{xx} (\sigma_{2 N_x m' + m}^x)^4 + \right. \\ 
	& \left. + \mathcal{Z}_{2 N_x m' + 2 N_x - m, 2N_y n'}^{xx} (\sigma_{2 N_x m' + 2 N_x - m}^x)^4 \right],
\end{split} \end{equation}	
for $n \neq 0$
\begin{equation} \begin{split}
	\tilde{E}_{0n} &= 2 \sum_{n'=0}^{\zeta_y-1} \sum_{m' = 0}^{\zeta_x-1} (\sigma_{2 N_x m'}^x)^4 \times \\
	& \times \left[ \mathcal{Z}_{2 N_x m', 2N_y n' + n}^{xx} (\sigma_{2N_y n' + n}^y)^2 + \right. \\
	& \left. + \mathcal{Z}_{2 N_x m', 2N_y n' + 2 N_y - n}^{xx} (\sigma_{2N_y n' + 2 N_y - n}^y)^2 \right],
\end{split} \end{equation}	
and for $m \neq 0, n \neq 0$
\begin{equation} \begin{split}
	& \tilde{E}_{mn} = \sum_{n'=0}^{\zeta_y-1} \sum_{m' = 0}^{\zeta_x-1} (\sigma_{2N_y n' + n}^y)^2 \times \\
	& \times \left[ \mathcal{Z}_{2 N_x m' + m, 2N_y n' + n}^{xx} (\sigma_{2 N_x m' + m}^x)^4 \right. + \\
	& \left. + \mathcal{Z}_{2 N_x m' + 2 N_x - m, 2N_y n' + n}^{xx} (\sigma_{2 N_x m' + 2 N_x - m}^x)^4 \right] + \\
	& + \sum_{n'=0}^{\zeta_y-1} \sum_{m' = 0}^{\zeta_x-1} (\sigma_{2N_y n' + 2 N_y - n}^y)^2 \times \\
	& \times \left[ \mathcal{Z}_{2 N_x m' + m, 2N_y n' + 2 N_y - n}^{xx} (\sigma_{2 N_x m' + m}^x)^4 \right. + \\
	& \left. + \mathcal{Z}_{2 N_x m' + 2 N_x - m, 2N_y n' + 2 N_y - n}^{xx} (\sigma_{2 N_x m' + 2 N_x - m}^x)^4 \right].
\end{split} \end{equation}
	
Equation (\ref{Z^xxFF_ik_FFT}) is a sum of two-dimensional FTs of the $2N_x \times 2N_y$ matrix $- (\Delta x \Delta y/4)^2 \tilde{E}$. Therefore $Z^{xxFF}_{ik}$ can be found by calling the two-dimensional inverse FFT: $E = \mathrm{FFT}_{2D}^{-1}(\tilde{E})$ and taking the corresponding elements of the resulting matrix: 
\begin{equation} \begin{split}
	Z^{xxFF}_{ik} &= - \frac{1}{16} (\Delta x \Delta y)^2 \times \\
	&\times \left[ E_{i_x + k_x, i_y + k_y + 1} - E_{i_x + k_x, i_y - k_y} + \right. \\
	&+ \left. E_{i_x - k_x, i_y + k_y + 1} - E_{i_x - k_x, i_y - k_y} \right],
	\label{Z^xxFF_ik_FFT2}
\end{split} \end{equation}
where $i_x, i_y$, $k_x, k_y$ are the mesh indices corresponding to the $i$-th and $k$-th basis functions.

\section{Numerical Examples}\label{Num_ex}

\begin{figure}[h]
	\centering
	\includegraphics[height=6cm]{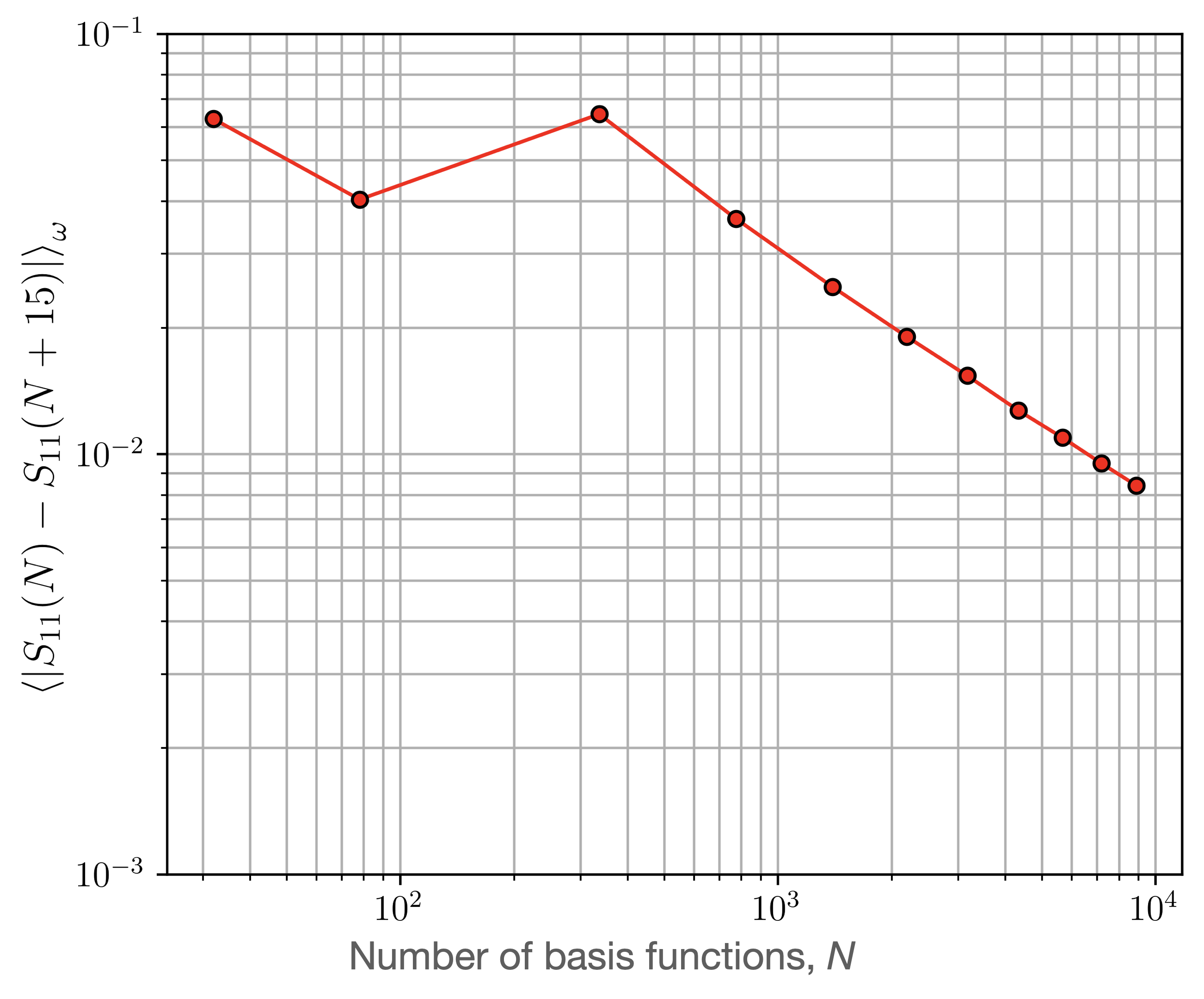} \includegraphics[height=6cm]{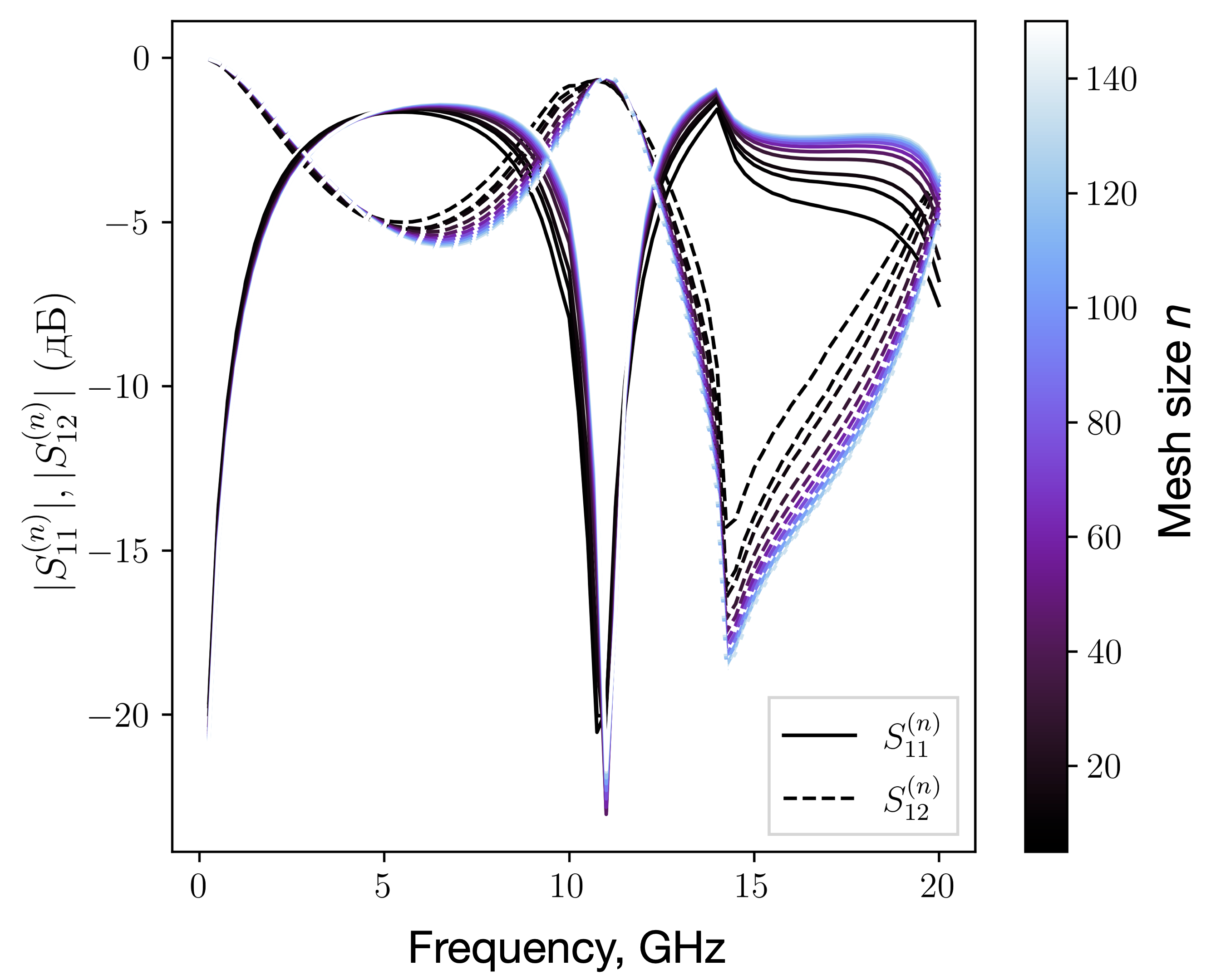}\\
	\caption{Top: An example of convergence of the $S_{11}$ parameter averaged over the spectrum for a metal strip in a waveguide as the number of basis functions increases. Bottom: the corresponding S-parameter spectrum for various numbers of basis functions.}
	\label{fig:convergence}
\end{figure}

It should be noted that the rooftop basis functions employed in this work, which are a specific type of piecewise-polynomial basis, facilitate the formulation of a fast, FFT-based method for calculating currents. However, they do not accurately approximate currents near the edges of metallic structures, where singularities occur (for instance, the $\sim 1/r^{1/2}$ singularity at the edge of an infinitely thin strip) \cite{mittra1971analytical}. However, due to their convenience in analytical calculations and the availability of efficient computational formulations, they are widely used in practice; they allow one to obtain integral characteristics of PCBs with accuracy sufficient for practical applications and electronic device design. It is also worth noting that some studies explore specialized basis functions tailored to specific types of singularities within the framework of MoM for planar structures \cite{vandenbosch2001general,hanninen2006singularity}. To demonstrate the applicability of the basis functions used here, Fig.~\ref{fig:convergence} shows the convergence of the $S_{11}$ parameter (averaged over the spectrum) and the corresponding S-parameter spectrum, calculated using varying numbers of basis functions for an infinitely thin, perfectly conducting strip within a waveguide. The calculation assumes a $15 \times 15$ mm waveguide containing a 3 mm wide strip that spans the entire cross-section of the waveguide, with ports located at either end. The plots indicate that the positions of the resonances are well-reproduced even with a small number of basis functions, and that, beyond a certain threshold, the convergence follows a clear polynomial trend.

Next we show more complex examples of S-parameter calculations for PCBs within shielded multilayer waveguides with perfectly conducting walls. It is assumed that the planar metallic elements have negligible thickness and zero resistance. The ports located on the enclosure walls have an impedance of 50 ohms. All S-parameter calculation results presented below were obtained without the use of a de-embedding procedure \cite{RAUTIO2014151}.

\begin{figure}[h]
	\centering
	\includegraphics[width=0.3\textwidth]{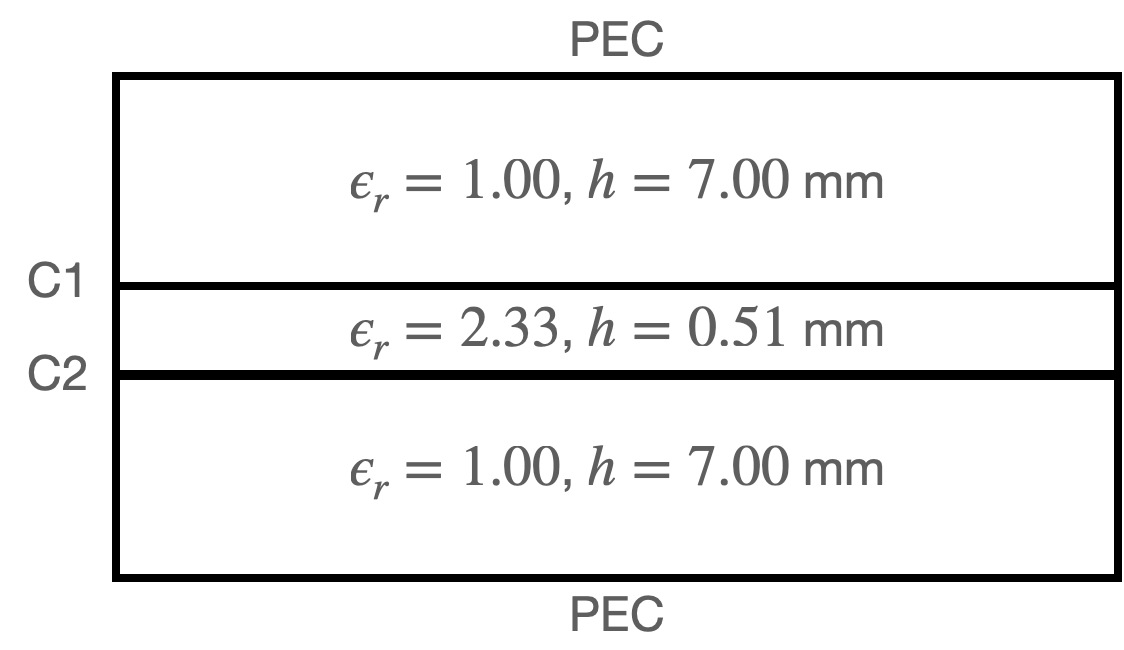}\\
	\vspace{0.1in}
	\includegraphics[width=0.3\textwidth]{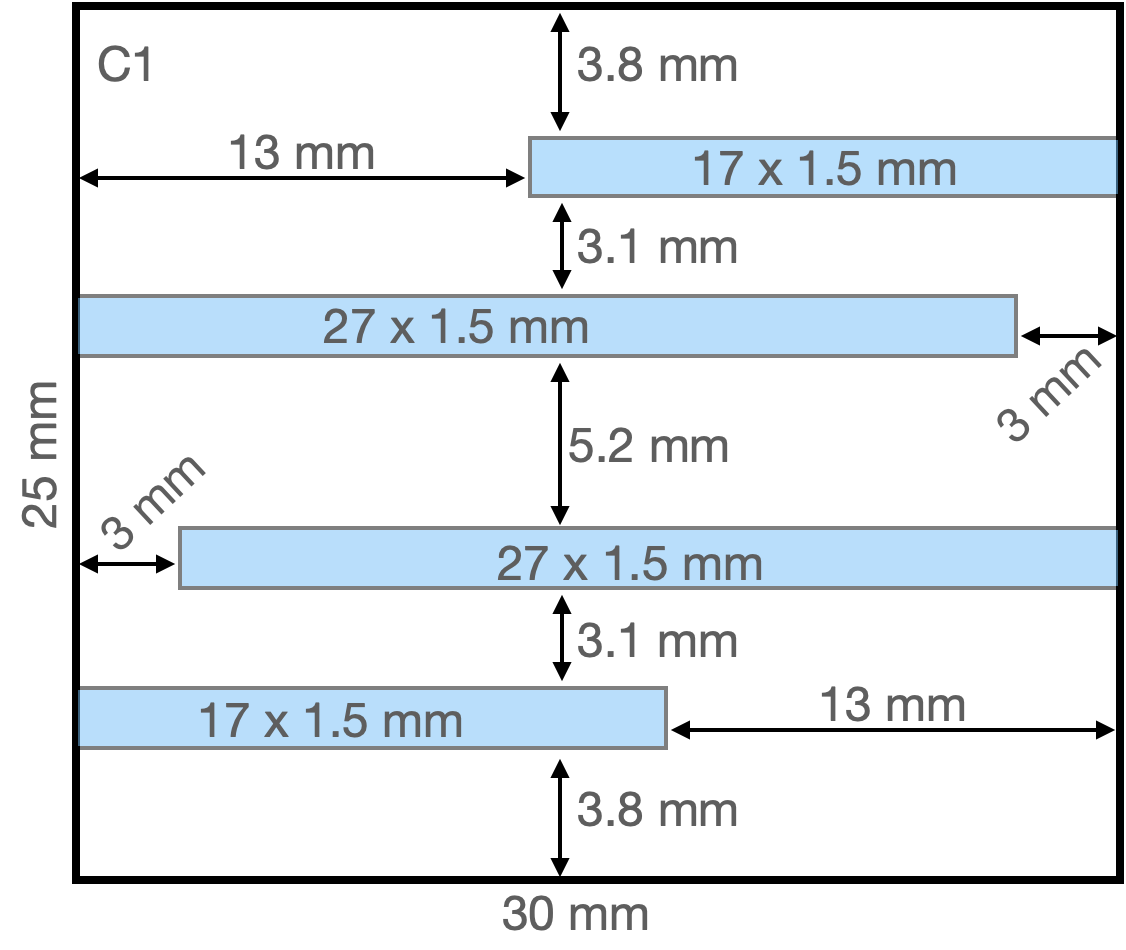}\\
	\vspace{0.1in}
	\includegraphics[width=0.3\textwidth]{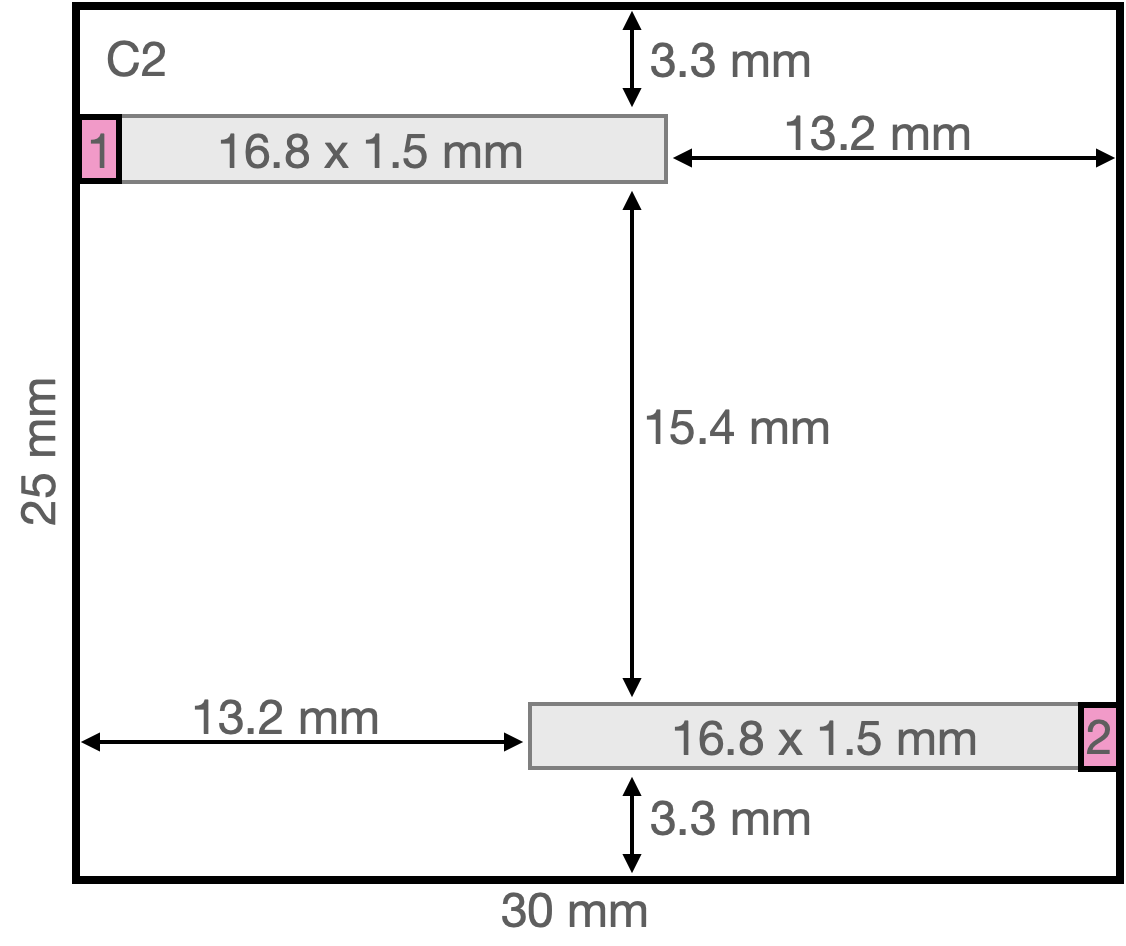}
	\caption{Filter with three layers and six metal strips: model geometry, dimensions and layer permittivities. Ports 1 and 2 are denoted by pink rectangles.}
	\label{fig:filter1_model}
\end{figure}

\begin{figure}[h]
	\centering
	\includegraphics[width=0.45\textwidth]{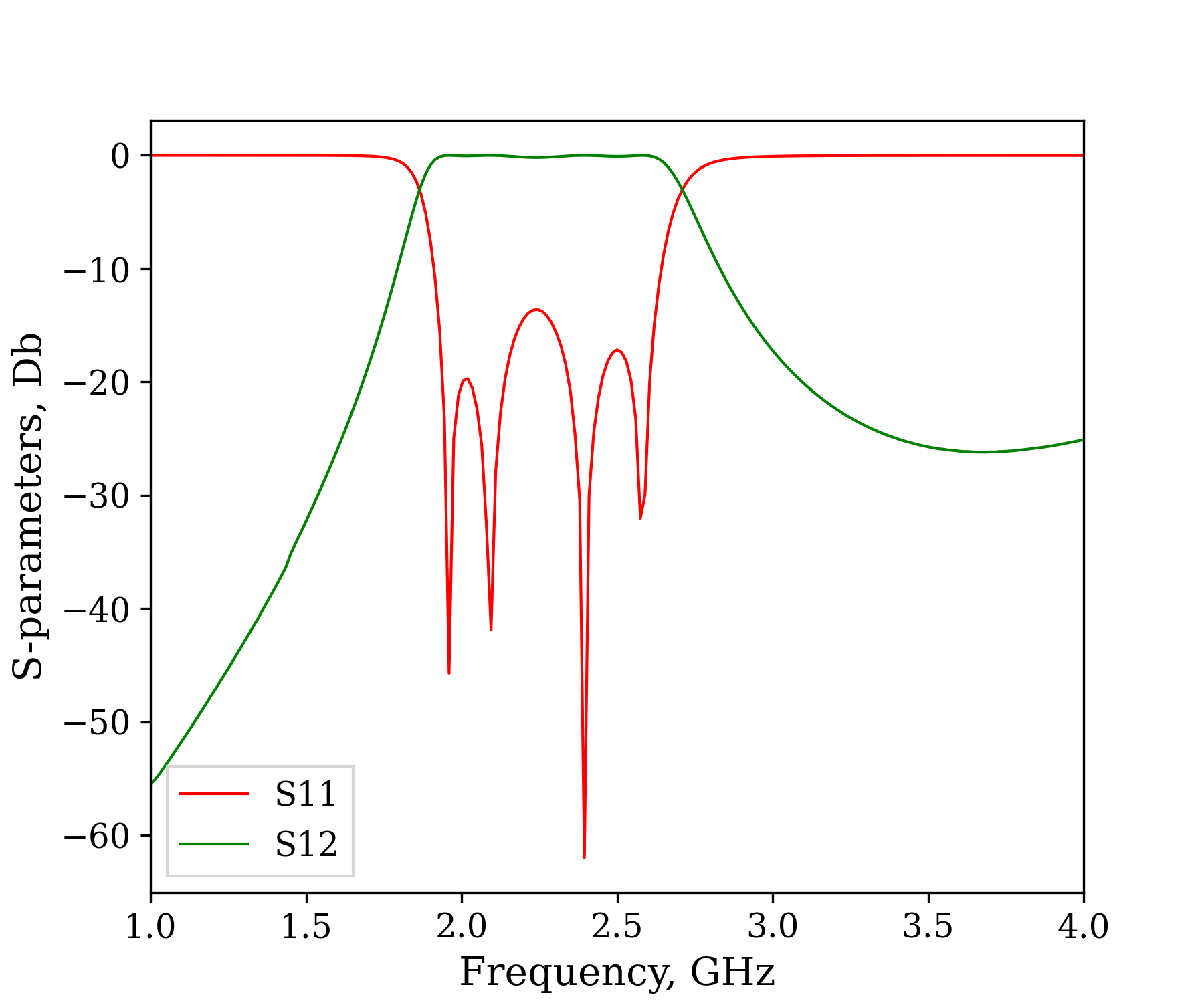}
	\caption{Filter with three layers and six metal strips: frequency dependence of $S_{11}$ and $S_{21}$ parameters.}
	\label{fig:filter1_Sparam}
\end{figure}

First, we consider a filter with three dielectric layers and two metallization layers in a waveguide section with the PEC boundary conditions at the top and bottom of the waveguide. The example is similar to \cite{Melcon1998}, Fig.~5.11, and the geometry of the problem is shown in Fig.~\ref{fig:filter1_model}. The upper metallization layer C1 consists of four strips, and the lower metallization layer constsis of two strips. Two delta-gap ports, representing the connection of coaxial cables through the waveguide walls, are attached to the strips at the C2 level. The simulated frequency range is from 1 to 4 GHz. Calculation was performed on a regular grid with the step of 0.1 mm ($300 \times 250$ elements). Figure \ref{fig:filter1_Sparam} shows frequency dependence of the scattering parameters $S_{11}$ and $S_{21}$. Our results agree well with the results in \cite{Melcon1998}, Fig.~5.11(b) if we take into account minor differences in the conductor geometry.

\begin{figure}[h]
	\centering
	\includegraphics[width=0.3\textwidth]{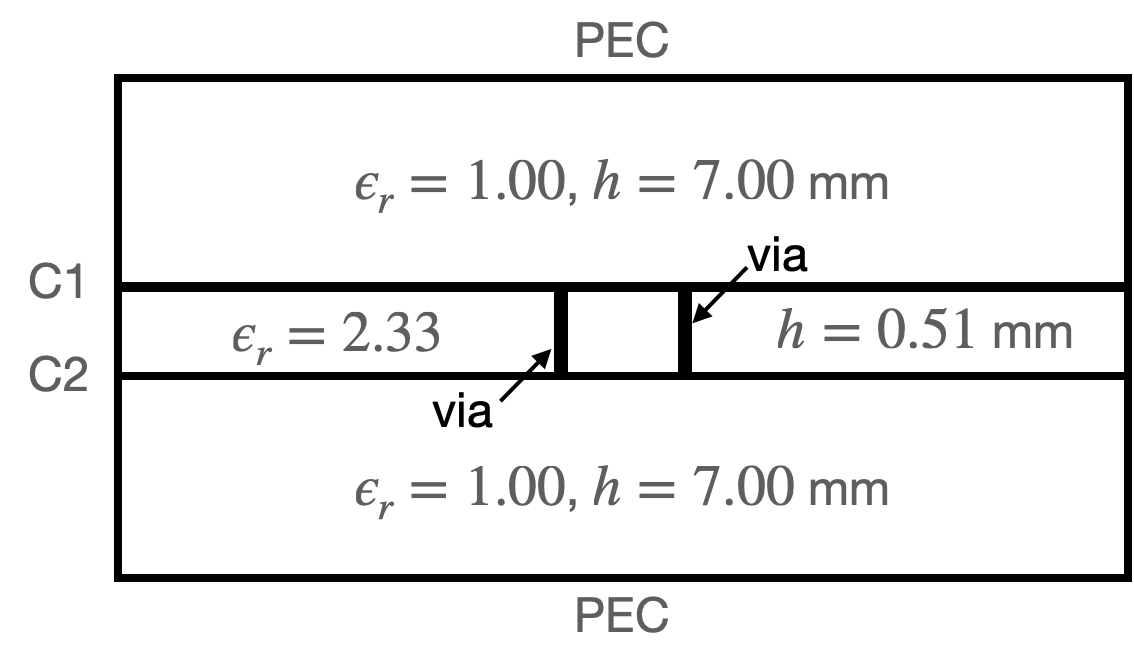}\\
	\vspace{0.1in}
	\includegraphics[width=0.3\textwidth]{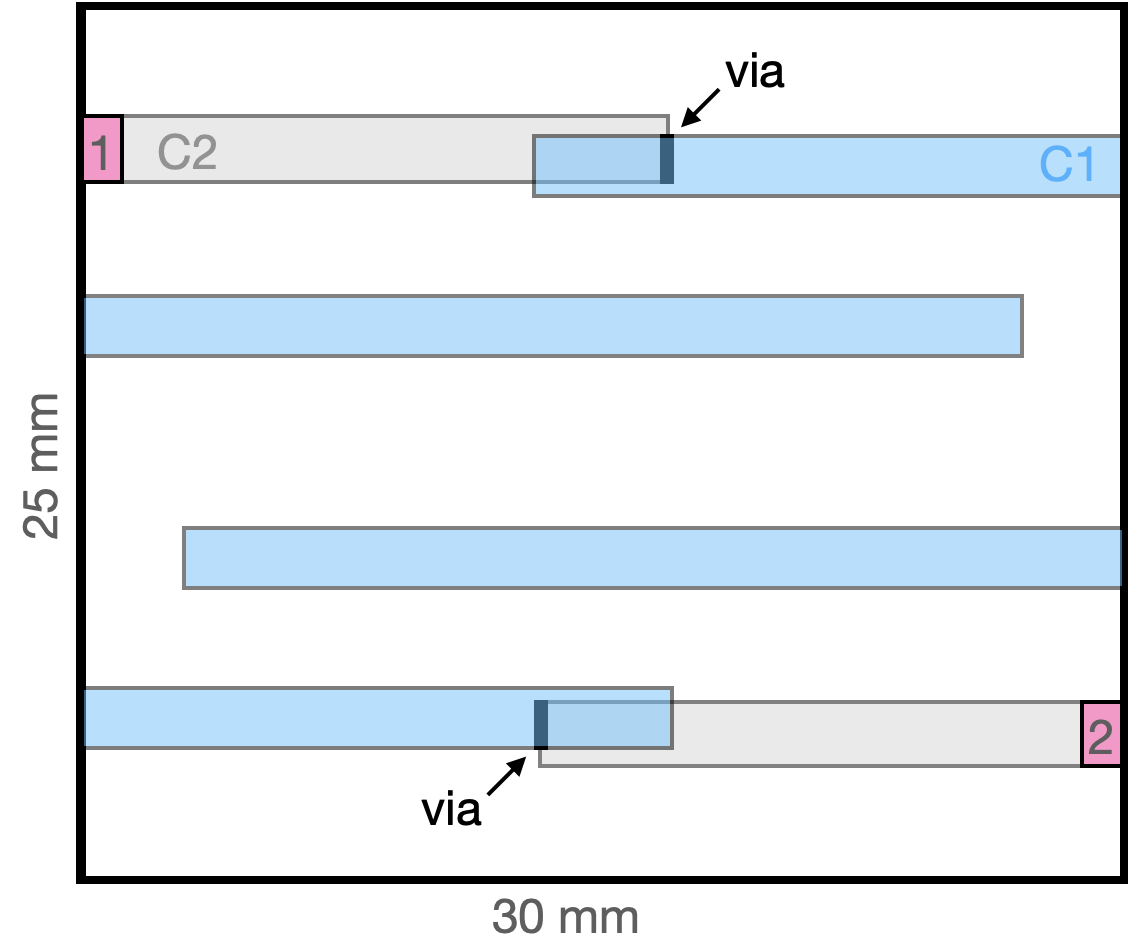}\\
	\vspace{0.1in}
	\includegraphics[width=0.3\textwidth]{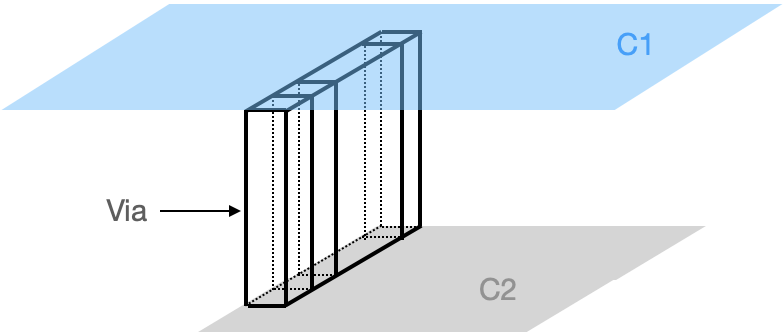}\\
	\vspace{0.1in}
	\caption{Filter with three layers and six metal strips $+$ interconnect: model geometry, dimensions and layer permittivities and interconnect zoom view. Interconnect is donoted by thick black lines. Ports 1 and 2 are denoted by pink rectangles.}
	\label{fig:filter2_model}
\end{figure}

\begin{figure}[h]
	\centering
	\includegraphics[width=0.45\textwidth]{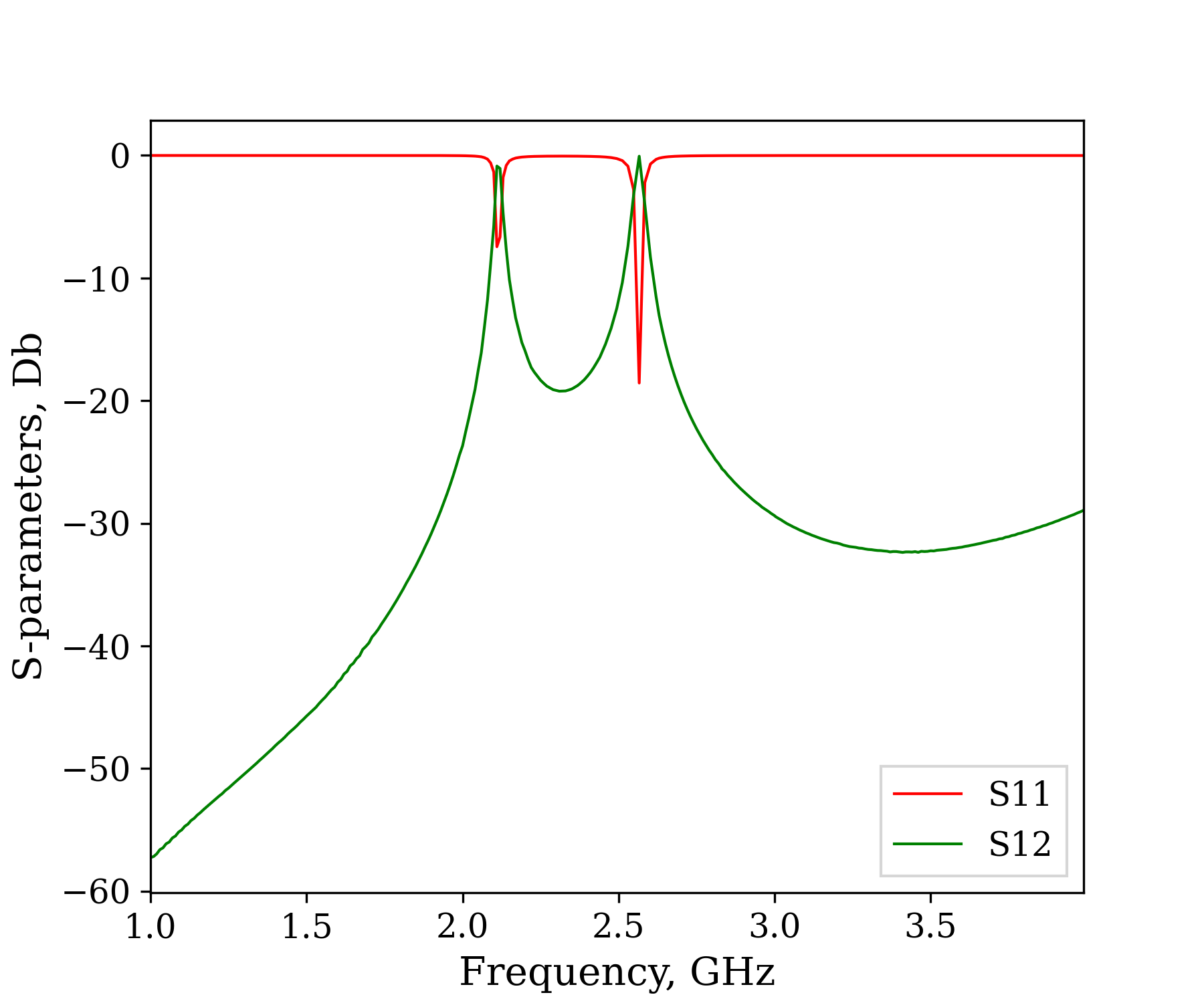}
	\caption{Filter with three layers, six metal strips and two interconnect sections: frequency dependence of $S_{11}$ and $S_{21}$ parameters.}
	\label{fig:filter2_Sparam}
\end{figure}

Next, we modify the considered filter with two vertical interconnects between the strips on level C1 and C2. The geometry is shown in Fig.~\ref{fig:filter2_model}. The via sections are denoted by thick black lines. They are modelled as single layers of volume basis functions placed side-by-side along the width of the strip overlap as shown in Fig.~\ref{fig:filter2_model}, bottom image. Fig.~\ref{fig:filter2_Sparam} shows the frequency dependence of the scattering parameters $S_{11}$ and $S_{21}$.

\section{Conclusion}\label{conc_sec}

In this paper, we have revisited the problem of shielded PCB modelling using the Method of Moments for electric field integral equation. We have shown how all components of the Green's function for arbitrary source and observation points in a layered rectangular waveguide can be explicitly expressed using the S-matrix method. A convenient framework is proposed for expressing the Green’s function components via three sets of S-matrices associated with the dielectric substrate layers. We have explicitly summarized the overlap integrals of our Green's function components with rooftop surface basis functions, as well as volume pulse and linear basis functions. A new derivation for the expansion of the MoM matrix elements is presented, facilitating computational acceleration via the Fast Fourier Transform (FFT). Two examples are included to demonstrate the method’s capabilities: the first is a well-known case from the literature, with the results obtained here showing strong agreement with the published data; the second is a modification of the first, designed to illustrate the potential impact of vertical interconnects. In summary, this work presents an S-matrix based formulation for modeling PCBs that is fully suitable for practical implementation. Furthermore, the Green’s function represented through these specific sets of S-matrices can be integrated with other types of basis functions (such as RWG basis functions) to model more complex geometries on triangular meshes within a multilayered waveguide.

\section{Acknowledgements}

The work was supported by the Russian Science Foundation, grant No~22-11-00153-$\Pi$.

\bibliographystyle{IEEEtran}
\bibliography{mom}

\begin{thebibliography}{10}
\providecommand{\url}[1]{#1}
\csname url@samestyle\endcsname
\providecommand{\newblock}{\relax}
\providecommand{\bibinfo}[2]{#2}
\providecommand{\BIBentrySTDinterwordspacing}{\spaceskip=0pt\relax}
\providecommand{\BIBentryALTinterwordstretchfactor}{4}
\providecommand{\BIBentryALTinterwordspacing}{\spaceskip=\fontdimen2\font plus
\BIBentryALTinterwordstretchfactor\fontdimen3\font minus
  \fontdimen4\font\relax}
\providecommand{\BIBforeignlanguage}[2]{{%
\expandafter\ifx\csname l@#1\endcsname\relax
\typeout{** WARNING: IEEEtran.bst: No hyphenation pattern has been}%
\typeout{** loaded for the language `#1'. Using the pattern for}%
\typeout{** the default language instead.}%
\else
\language=\csname l@#1\endcsname
\fi
#2}}
\providecommand{\BIBdecl}{\relax}
\BIBdecl

\bibitem{Chandler1995}
\BIBentryALTinterwordspacing
N.~Chandler and S.~Tyler, ``Ultra-fine feature printed circuits and multi-chip
  modules,'' \emph{Microelectronics Journal}, vol.~26, no.~4, pp. 393--404,
  1995. [Online]. Available:
  \url{https://www.sciencedirect.com/science/article/pii/002626929598940S}
\BIBentrySTDinterwordspacing

\bibitem{Walter1996}
\BIBentryALTinterwordspacing
P.~Walter, ``{Bond testing enters mainstream PCB assembly},''
  \emph{Microelectronics Journal}, vol.~27, no.~1, pp. i--v, 1996. [Online].
  Available:
  \url{https://www.sciencedirect.com/science/article/pii/S002626929690018X}
\BIBentrySTDinterwordspacing

\bibitem{Prentice1997}
\BIBentryALTinterwordspacing
T.~Prentice, ``{Dispensing answers the call: Speed, accuracy, and flexibility
  meet PCB assembly needs},'' \emph{Microelectronics Journal}, vol.~28, no.~1,
  pp. vii--xi, 1997. [Online]. Available:
  \url{https://www.sciencedirect.com/science/article/pii/S002626929787869X}
\BIBentrySTDinterwordspacing

\bibitem{Morrissey1998}
\BIBentryALTinterwordspacing
A.~Morrissey, G.~Kelly, J.~Alderman, J.~Barrett, C.~Lyden, and L.~O'Rourke,
  ``{Some issues for microsystem packaging in plastic and 3D},''
  \emph{Microelectronics Journal}, vol.~29, no.~9, pp. 645--650, 1998, low
  Dimensional Structures and Devices: Micromachined Devices. [Online].
  Available:
  \url{https://www.sciencedirect.com/science/article/pii/S0026269298000299}
\BIBentrySTDinterwordspacing

\bibitem{Bai2025}
\BIBentryALTinterwordspacing
X.~Bai, Y.~Wu, S.~Zhen, Z.~Gao, and W.~Wang, ``{A 8-10 GHz compact low noise
  amplifier MMIC with high linearity based on GaAs technology},''
  \emph{Microelectronics Journal}, vol. 162, p. 106742, 2025. [Online].
  Available:
  \url{https://www.sciencedirect.com/science/article/pii/S1879239125001912}
\BIBentrySTDinterwordspacing

\bibitem{Zeng2025}
\BIBentryALTinterwordspacing
Y.~Zeng, Y.~Wu, Y.~Yang, L.~Pan, and W.~Wang, ``{A novel minimized
  millimeter-wave on-chip spoof surface plasmon polariton and its applications
  based on IPD technology},'' \emph{Microelectronics Journal}, vol. 162, p.
  106755, 2025. [Online]. Available:
  \url{https://www.sciencedirect.com/science/article/pii/S1879239125002048}
\BIBentrySTDinterwordspacing

\bibitem{Rautio2007}
B.~J. Rautio, ``{Applications of advanced shielded planar EM analysis
  techniques to antenna analysis},'' \emph{2nd European Conference on Antennas
  and Propagation (EuCAP 2007) Proceedings}, 2007.

\bibitem{Gibson2008}
W.~C. Gibson, \emph{{The Method of Moments in electromagnetics}}.\hskip 1em
  plus 0.5em minus 0.4em\relax Boca Raton, FL: Chapman \& Hall/CRC, 2008.

\bibitem{Swanson2003-fi}
D.~G. Swanson~Jr and W.~J.~R. Hoefer, \emph{\BIBforeignlanguage{en}{Microwave
  circuit modeling using electromagnetic field simulation}}, ser. Microwave
  Library.\hskip 1em plus 0.5em minus 0.4em\relax Norwood, MA: Artech House,
  May 2003.

\bibitem{Lancellotti2022}
V.~Lancellotti, ``{Volume 2: Field representations and the Method of
  Moments},'' in \emph{Advanced Theoretical and Numerical
  Electromagnetics}.\hskip 1em plus 0.5em minus 0.4em\relax London, United
  Kingdom: SciTech Publishing, 2022.

\bibitem{Okhmatovski2024}
V.~Okhmatovski and S.~Zheng, \emph{{Theory and Computation of Electromagnetic
  Fields in Layered Media}}.\hskip 1em plus 0.5em minus 0.4em\relax Wiley-IEEE
  Press, 2024.

\bibitem{Khalil1999}
A.~I. Khalil, A.~B. Yakovlev, and M.~B. Steer, ``{Efficient Method-of-Moments
  formulation for the modeling of planar conductive layers in a shielded
  guided-wave structure},'' \emph{IEEE Transactions on Microwave Theory and
  Techniques}, vol.~47, no.~9, pp. 1730--1736, 1999.

\bibitem{Davidson2005}
D.~B. Davidson, \emph{{Computational Electromagnetics for RF and Microwave
  Engineering}}.\hskip 1em plus 0.5em minus 0.4em\relax Cambridge University
  Press, 2005.

\bibitem{Crespo-Valero2006}
P.~Crespo-Valero, M.~Mattes, I.~Stevanovi{\'{c}}, and J.~R. Mosig, ``{Analysis
  of multilayer boxed printed circuits},'' \emph{Proceedings of the
  Mediterranean Electrotechnical Conference - MELECON}, vol. 2006, no.~4, pp.
  206--209, 2006.

\bibitem{Abdul-Gaffoor2000}
M.~R. Abdul-Gaffoor, ``{Simple and efficient full wave analysis of
  electromagnetic coupling in multilayer printed circuit board layouts},''
  Ph.D. dissertation, University of Mississippi, 2000.

\bibitem{Abdul-Gaffoor2002}
M.~R. Abdul-Gaffoor, H.~K. Smith, A.~A. Kishk, and A.~W. Glisson, ``{Simple and
  efficient full-wave modeling of electromagnetic coupling in realistic RF
  multilayer PCB layouts},'' \emph{IEEE Transactions on Microwave Theory and
  Techniques}, vol.~50, no.~6, pp. 1445--1457, 2002.

\bibitem{Michalski1997}
K.~A. Michalski and J.~R. Mosig, ``{Multilayered media Green's functions in
  integral equation formulations},'' \emph{IEEE Transactions on Antennas and
  Propagation}, vol.~45, no.~3, pp. 508--519, 1997.

\bibitem{Kinayman2005}
N.~Kinayman and M.~I. Aksun, \emph{{Modern Microwave Circuits}}.\hskip 1em plus
  0.5em minus 0.4em\relax Artech House, 2005.

\bibitem{Okhmatovski2009}
V.~Okhmatovski, M.~Yuan, I.~Jeffrey, and R.~Phelps, ``{A three-dimensional
  precorrected FFT algorithm for fast method of moments solutions of the
  mixed-potential integral equation in layered media},'' \emph{IEEE
  Transactions on Microwave Theory and Techniques}, vol.~57, no.~12, pp.
  3505--3517, 2009.

\bibitem{Chew2001}
W.~Chew, E.~Michielssen, J.~Song, and J.~Jin, \emph{{Fast and Efficient
  Algorithms in Computational Electromagnetics}}.\hskip 1em plus 0.5em minus
  0.4em\relax Artech House, 2001.

\bibitem{YlaOijala2003}
P.~Yla-Oijala and M.~Taskinen, ``{Calculation of CFIE impedance matrix elements
  with RWG and n/spl times/RWG functions},'' \emph{IEEE Transactions on
  Antennas and Propagation}, vol.~51, no.~8, pp. 1837--1846, 2003.

\bibitem{Kinayman1995}
N.~Kinayman and M.~I. A., ``Comparative study of acceleration techniques for
  integrals and series in electromagnetic problems,'' \emph{Radio Science},
  vol.~30, no.~6, pp. 1713--1722, 1995.

\bibitem{Hashemi1995}
S.~Hashemi-Yeganeh, ``On the summation of double infinite series field
  computations inside rectangular cavities,'' \emph{IEEE Transactions on
  Microwave Theory and Techniques}, vol.~43, no.~3, pp. 641--646, 1995.

\bibitem{Eleftheriades1996a}
G.~Eleftheriades, J.~Mosig, and M.~Guglielmi, ``A fast integral equation
  technique for shielded planar circuits defined on nonuniform meshes,''
  \emph{IEEE Transactions on Microwave Theory and Techniques}, vol.~44, no.~12,
  pp. 2293--2296, 1996.

\bibitem{Nguyen2015}
H.~Q. Nguyen and M.~Unser, ``{Generalized Poisson summation formula for
  tempered distributions},'' in \emph{2015 International Conference on Sampling
  Theory and Applications (SampTA)}, 2015, pp. 1--5.

\bibitem{Park1997}
S.-O. Park and C.~Balanis, ``{Analytical technique to evaluate the asymptotic
  part of the impedance matrix of Sommerfeld-type integrals},'' \emph{IEEE
  Transactions on Antennas and Propagation}, vol.~45, no.~5, pp. 798--805,
  1997.

\bibitem{Soler2008}
F.~J.~P. Soler, F.~D.~Q. Pereira, D.~C. Rebenaque, A.~A. Melcon, and J.~R.
  Mosig, ``{A novel efficient technique for the calculation of the Green's
  functions in rectangular waveguides based on accelerated series
  decomposition},'' \emph{IEEE Transactions on Antennas and Propagation},
  vol.~56, no.~10, pp. 3260--3270, 2008.

\bibitem{delaCruz2023}
A.~M.~H. de~la Cruz, C.~G. Molina, F.~D. Quesada~Pereira, A.~A. Melcon, and
  V.~E.~B. Esbert, ``{Efficient calculation of the 3-D rectangular waveguide
  Green’s functions derivatives by the Ewald method},'' \emph{IEEE
  Transactions on Microwave Theory and Techniques}, vol.~71, no.~12, pp.
  5171--5181, 2023.

\bibitem{Rautio1986}
J.~C. Rautio, ``{A time-harmonic electromagnetic analysis of shielded
  microstrip circuits},'' Ph.D. dissertation, Syracuse University, 1986.

\bibitem{Hill1991}
A.~Hill and V.~Tripathi, ``An efficient algorithm for the three-dimensional
  analysis of passive microstrip components and discontinuities for microwave
  and millimeter-wave integrated circuits,'' \emph{IEEE Transactions on
  Microwave Theory and Techniques}, vol.~39, no.~1, pp. 83--91, 1991.

\bibitem{Rautio2014a}
B.~J. Rautio, V.~I. Okhmatovski, A.~C. Cangellaris, J.~C. Rautio, and J.~K.
  Lee, ``{The Unified-FFT algorithm for fast electromagnetic analysis of planar
  integrated circuits printed on layered media inside a rectangular
  enclosure},'' \emph{IEEE Transactions on Microwave Theory and Techniques},
  vol.~62, no.~5, pp. 1112--1121, 2014.

\bibitem{Rautio2015}
B.~Rautio, V.~I. Okhmatovski, and J.~K. Lee, ``{The unified-FFT grid totalizing
  algorithm for fast O(N logN) method of moments electromagnetic analysis with
  accuracy to machine precision},'' \emph{Progress in Electromagnetics
  Research}, vol. 154, no. November, pp. 101--114, 2015.

\bibitem{Strang1986}
G.~Strang, ``{A Proposal for Toeplitz matrix calculations},'' \emph{Studies in
  Applied Mathematics}, vol.~74, no.~2, pp. 171--176, 1986.

\bibitem{Strang1989}
------, ``{Toeplitz equations by conjugate gradients with circulant
  preconditioner},'' \emph{SIAM Journal on Scientific and Statistical
  Computing}, vol.~10, no.~1, pp. 104--119, 1989.

\bibitem{Zhao1998}
J.-S. Zhao, W.~C. Chew, C.-C. Lu, E.~Michielssen, and J.~Song,
  ``Thin-stratified medium fast-multipole algorithm for solving microstrip
  structures,'' \emph{IEEE Transactions on Microwave Theory and Techniques},
  vol.~46, no.~4, pp. 395--403, 1998.

\bibitem{Golestanirad2010}
\BIBentryALTinterwordspacing
L.~Golestanirad, M.~Mattes, and J.~R. Mosig, ``{On the application of symmetry
  conditions and the convergence rate of modal series in the MOM-based integral
  equation analysis of laterally shielded multilayered media},''
  \emph{Microwave and Optical Technology Letters}, vol.~52, no.~1, pp.
  221--226, 2010. [Online]. Available:
  \url{https://onlinelibrary.wiley.com/doi/abs/10.1002/mop.24868}
\BIBentrySTDinterwordspacing

\bibitem{Golestanirad2010a}
------, ``An efficient technique to accelerate the simulation of passive
  shielded microwave circuits,'' \emph{Proceedings of the Fourth European
  Conference on Antennas and Propagation}, pp. 1--4, 2010.

\bibitem{Rautio1987}
J.~C. Rautio and R.~F. Harrington, ``{An electromagnetic time-harmonic analysis
  of shielded microstrip circuits},'' pp. 726--730, 1987.

\bibitem{Rautio2005}
\BIBentryALTinterwordspacing
J.~C. Rautio, \emph{Applied Numerical Electromagnetic Analysis for Planar High
  Frequency Circuits}.\hskip 1em plus 0.5em minus 0.4em\relax John Wiley and
  Sons, Ltd, 2005. [Online]. Available:
  \url{https://onlinelibrary.wiley.com/doi/abs/10.1002/0471654507.eme040}
\BIBentrySTDinterwordspacing

\bibitem{Melcon1999}
A.~Melcon, J.~Mosig, and M.~Guglielmi, ``{Efficient CAD of boxed microwave
  circuits based on arbitrary rectangular elements},'' \emph{IEEE Transactions
  on Microwave Theory and Techniques}, vol.~47, no.~7, pp. 1045--1058, 1999.

\bibitem{Stevanovic2009}
I.~Stevanovic, F.~Merli, P.~Crespo-Valero, W.~Simon, S.~Holzwarth, M.~Mattes,
  and J.~R. Mosig, ``Integral equation modeling of waveguide-fed planar
  antennas,'' \emph{IEEE Antennas and Propagation Magazine}, vol.~51, no.~6,
  pp. 82--92, 2009.

\bibitem{Michalski2019}
\BIBentryALTinterwordspacing
K.~A. Michalski, ``Modal transmission line theory of plane wave excited layered
  media with multiple conductive anisotropic sheets at the interfaces,''
  \emph{Journal of Quantitative Spectroscopy and Radiative Transfer}, vol. 226,
  pp. 19--28, 2019. [Online]. Available:
  \url{https://www.sciencedirect.com/science/article/pii/S0022407318308987}
\BIBentrySTDinterwordspacing

\bibitem{Khalil1999a}
A.~I. Khalil and M.~B. Steer, ``{A generalized scattering matrix method using
  the Method of Moments for electromagnetic analysis of multilayered structures
  in waveguide},'' \emph{IEEE Transactions on Microwave Theory and Techniques},
  vol.~47, no.~9, pp. 2151--2157, 1999.

\bibitem{Yakovlev1999}
A.~Yakovlev, A.~Khalil, C.~Hicks, and M.~Steer, ``Electromagnetic modeling of a
  waveguide-based strip-to-slot transition module for application to spatial
  power combining systems,'' in \emph{IEEE Antennas and Propagation Society
  International Symposium. 1999 Digest. Held in conjunction with: USNC/URSI
  National Radio Science Meeting (Cat. No.99CH37010)}, vol.~1, 1999, pp.
  286--289 vol.1.

\bibitem{Yakovlev2000}
A.~Yakovlev, A.~Khalil, C.~Hicks, A.~Mortazawi, and M.~Steer, ``The generalized
  scattering matrix of closely spaced strip and slot layers in waveguide,''
  \emph{IEEE Transactions on Microwave Theory and Techniques}, vol.~48, no.~1,
  pp. 126--137, 2000.

\bibitem{Yakovlev2002}
A.~Yakovlev, S.~Ortiz, M.~Ozkar, A.~Mortazawi, and M.~Steer, ``{Electric dyadic
  Green’s functions for modeling resonance and coupling effects in
  waveguide-based aperture-coupled patch arrays},'' \emph{Aces Journal},
  vol.~17, no.~2, pp. 123--133, 2002.

\bibitem{Li1997}
L.-W. Li, P.-S. Kooi, M.-S. Leong, T.-S. Yeo, and S.-L. Ho, ``{Input impedance
  of a probe-excited semi-infinite rectangular waveguide with arbitrary
  multilayered loads: Part II - A full-wave analysis},'' \emph{IEEE
  Transactions on Microwave Theory and Techniques}, vol.~45, no.~3, pp.
  321--329, 1997.

\bibitem{Molina2020}
C.~Gomez~Molina, F.~Q. Pereira, A.~A. Melcon, S.~Marini, M.~A. Sanchez-Soriano,
  V.~E. Boria, and M.~Guglielmi, ``Multimode equivalent network for boxed
  multilayer arbitrary planar circuits,'' \emph{IEEE Transactions on Microwave
  Theory and Techniques}, vol.~68, no.~7, pp. 2501--2514, 2020.

\bibitem{Ohira2005}
M.~Ohira, H.~Deguchi, M.~Tsuji, and H.~Shigesawa, ``Novel waveguide filters
  with multiple attenuation poles using dual-behavior resonance of
  frequency-selective surfaces,'' \emph{IEEE Transactions on Microwave Theory
  and Techniques}, vol.~53, no.~11, pp. 3320--3326, 2005.

\bibitem{Stevanovic2004}
I.~Stevanovic and J.~Mosig, ``Efficient electromagnetic analysis of line-fed
  aperture antennas in thick conducting screens,'' \emph{IEEE Transactions on
  Antennas and Propagation}, vol.~52, no.~11, pp. 2896--2903, 2004.

\bibitem{Stevanovic2006}
I.~Stevanovic, P.~Crespo-Valero, and J.~Mosig, ``An integral-equation technique
  for solving thick irises in rectangular waveguides,'' \emph{IEEE Transactions
  on Microwave Theory and Techniques}, vol.~54, no.~1, pp. 189--197, 2006.

\bibitem{Rautio2014}
B.~J. Rautio, ``{The unified-FFT method for fast solution of integral equations
  as applied to shielded-domain electromagnetics},'' Ph.D. dissertation,
  Syracuse University, 1986.

\bibitem{Li2021}
\BIBentryALTinterwordspacing
N.~Li and M.~Miao, ``{Design of EMI and suppression structure based on
  bar-via},'' \emph{Microelectronics Journal}, vol. 112, p. 105049, 2021.
  [Online]. Available:
  \url{https://www.sciencedirect.com/science/article/pii/S0026269221000604}
\BIBentrySTDinterwordspacing

\bibitem{Rautio2019}
J.~C. Rautio and M.~Thelen, ``{Volume rooftop basis functions in shielded
  layered media},'' \emph{2019 IEEE MTT-S International Conference on Numerical
  Electromagnetic and Multiphysics Modeling and Optimization, NEMO 2019},
  vol.~c, pp. 1--4, 2019.

\bibitem{Rautio2020}
------, ``{A volume current based Method of Moments analysis of shielded planar
  3-D circuits in layered media},'' in \emph{2020 IEEE/MTT-S International
  Microwave Symposium (IMS)}, 2020, pp. 146--149.

\bibitem{Rautio2021}
J.~C. Rautio and M.~A. Thelen, ``{Method of Moments analysis of arbitrary
  structures in shielded layered media},'' \emph{IEEE Transactions on Microwave
  Theory and Techniques}, vol.~69, no.~1, pp. 509--517, 2021.

\bibitem{Nikellis2004}
K.~Nikellis, Y.~Koutsoyannopoulos, S.~Bantas, and N.~Uzunoglu, ``A fast
  full-wave modeling methodology for stripline structures with vertical
  interconnects in multi-layer dielectrics,'' in \emph{2004 Proceedings. 54th
  Electronic Components and Technology Conference (IEEE Cat. No.04CH37546)},
  vol.~1, 2004, pp. 225--230 Vol.1.

\bibitem{Nikellis2006}
K.~Nikellis, N.~Uzunoglu, Y.~Koutsoyannopoulos, and S.~Bantas, ``{Full-wave
  modeling of stripline structures in multilayer dielectrics},'' \emph{Progress
  in Electromagnetic Research}, vol.~57, pp. 253--264, 2006.

\bibitem{Aroudaki1994}
H.~Aroudaki and V.~Hansen, ``{Full-wave analysis of three-dimensional
  microstrip discontinuities by a new spectral domain approach},'' \emph{IEEE
  AP-S International Symposium, Proceedings}, vol.~3, p. 1694–1697, 1994.

\bibitem{Sonnet}
\BIBentryALTinterwordspacing
{Sonnet Software Inc.} Sonnet suites. [Online]. Available:
  \url{https://www.sonnetsoftware.com/}
\BIBentrySTDinterwordspacing

\bibitem{Sercu1994}
J.~Sercu, N.~Fache, and P.~Lagasse, ``First-order rooftop functions for the
  current discretisation in the method of moments solution of planar
  structures,'' in \emph{Proceedings of IEEE Antennas and Propagation Society
  International Symposium and URSI National Radio Science Meeting}, vol.~3,
  1994, pp. 2170--2173 vol.3.

\bibitem{Rao1982}
S.~M. Rao, D.~R. Wilton, and A.~W. Glisson, ``{Electromagnetic scattering by
  surfaces of arbitrary shape},'' \emph{IEEE Transactions on Antennas and
  Propagation}, vol.~30, no.~3, pp. 409--418, 1982.

\bibitem{Ko1988}
\BIBentryALTinterwordspacing
D.~Y.~K. Ko and J.~R. Sambles, ``Scattering matrix method for propagation of
  radiation in stratified media: attenuated total reflection studies of liquid
  crystals,'' \emph{J. Opt. Soc. Am. A}, vol.~5, no.~11, pp. 1863--1866, Nov
  1988. [Online]. Available:
  \url{https://opg.optica.org/josaa/abstract.cfm?URI=josaa-5-11-1863}
\BIBentrySTDinterwordspacing

\bibitem{Li1996}
\BIBentryALTinterwordspacing
L.~Li, ``Formulation and comparison of two recursive matrix algorithms for
  modeling layered diffraction gratings,'' \emph{J. Opt. Soc. Am. A}, vol.~13,
  no.~5, pp. 1024--1035, May 1996. [Online]. Available:
  \url{https://opg.optica.org/josaa/abstract.cfm?URI=josaa-13-5-1024}
\BIBentrySTDinterwordspacing

\bibitem{Volakis2012_iem}
\BIBentryALTinterwordspacing
J.~L. Volakis and K.~Sertel, \emph{Integral Equation Methods for
  Electromagnetics}.\hskip 1em plus 0.5em minus 0.4em\relax Institution of
  Engineering and Technology, Jan. 2012. [Online]. Available:
  \url{http://dx.doi.org/10.1049/SBEW045E}
\BIBentrySTDinterwordspacing

\bibitem{leontovich1948}
M.~A. Leontovich, ``\BIBforeignlanguage{russian}{On approximate boundary
  conditions for the electromagnetic field at the surface of highly conducting
  bodies},'' in \emph{\BIBforeignlanguage{russian}{Research on radio wave
  propagation. V. 2 (in Russian)}}.\hskip 1em plus 0.5em minus 0.4em\relax
  Publishing House of the USSR Academy of Sciences, 1948, pp. 5--12.

\bibitem{rytov1940skin}
C.~M. Rytov, ``\BIBforeignlanguage{russian}{Calculation of the skin effect
  using the perturbation method},'' \emph{\BIBforeignlanguage{russian}{Journal
  of Experimental and Theoretical Physics (in Russian)}}, vol.~10, no.~2, pp.
  180--189, 1940.

\bibitem{MarcuvitzWH1986}
\BIBentryALTinterwordspacing
N.~Marcuvitz, \emph{Waveguide Handbook}.\hskip 1em plus 0.5em minus 0.4em\relax
  The Institution of Engineering and Technology, 1986. [Online]. Available:
  \url{https://digital-library.theiet.org/doi/abs/10.1049/PBEW021E}
\BIBentrySTDinterwordspacing

\bibitem{redheffer1962relation}
R.~Redheffer, ``On the relation of transmission-line theory to scattering and
  transfer,'' \emph{Journal of Mathematics and Physics}, vol.~41, no. 1-4, pp.
  1--41, 1962.

\bibitem{Rautio2020b}
J.~C. Rautio, \emph{{Method of Moments Analysis of Structures Embedded in
  Shielded Layered Media}}, USA, 2020.

\bibitem{mittra1971analytical}
R.~Mittra, \emph{Analytical techniques in the theory of guided waves}.\hskip
  1em plus 0.5em minus 0.4em\relax The Macmillan Company, 1971.

\bibitem{vandenbosch2001general}
G.~A. Vandenbosch, ``General technique to calculate subsectional edge expansion
  functions (seefs) for microstrip antennas in arbitrary multilayered
  structures,'' in \emph{2001 Eleventh International Conference on Antennas and
  Propagation,(IEE Conf. Publ. No. 480)}, vol.~2.\hskip 1em plus 0.5em minus
  0.4em\relax IET, 2001, pp. 745--749.

\bibitem{hanninen2006singularity}
I.~Hanninen, M.~Taskinen, and J.~Sarvas, ``Singularity subtraction integral
  formulae for surface integral equations with rwg, rooftop and hybrid basis
  functions,'' \emph{Progress In Electromagnetics Research}, vol.~63, pp.
  243--278, 2006.

\bibitem{RAUTIO2014151}
J.~C. Rautio, ``Chapter 4 - high-frequency and microwave electromagnetic
  analysis calibration and de-embedding,'' in \emph{Microwave De-embedding},
  G.~Crupi and D.~M.-P. Schreurs, Eds.\hskip 1em plus 0.5em minus 0.4em\relax
  Oxford: Academic Press, 2014, pp. 151--188.

\bibitem{Melcon1998}
A.~A. Melcon, ``{Applications of the integral equation technique to the
  analysis and sinthesis of multilayered printed shielded microwave circuits
  and cavity backed antennas},'' Ph.D. dissertation, Universidad Polit´ecnica
  de Madrid, 1998.

\end{thebibliography}

\end{document}